\documentclass[aps,prb,reprint,superscriptaddress,twocolumn]{revtex4-2}
\usepackage[utf8]{inputenc}

\usepackage{amsmath}
\usepackage{graphicx}
\usepackage{lipsum}
\usepackage{physics}  %
\usepackage{verbatim}
\usepackage{CJKutf8}
\usepackage{wasysym}
\usepackage{xcolor}
\usepackage{hyperref}
\usepackage[normalem]{ulem}
\usepackage{microtype}

\newcommand{\lsi}{LSI, CNRS, CEA/DRF/IRAMIS, \'Ecole Polytechnique, Institut Polytechnique de Paris, F-91120 Palaiseau, France}
\newcommand{\etsf}{European Theoretical Spectroscopy Facility (ETSF)}
\newcommand{\ccq}{Center for Computational Quantum Physics, Flatiron Institute, 162 5th Avenue, New York, New York, USA}
\newcommand{\hofstra}{Department of Physics and Astronomy, Hofstra University, Hempstead, New York 11549, USA}
\newcommand{\soleil}{Synchrotron SOLEIL, L'Orme des Merisiers, Saint-Aubin, BP 48, F-91192 Gif-sur-Yvette, France}

\begin{document}

\title{Static Response of the Strongly Coupled Electron Gas from Diffusion Monte Carlo}

\author{Muhammed H. Güneş}
\affiliation{\lsi}%
\affiliation{\etsf}%

\author{Yubo Yang \begin{CJK}{UTF8}{gbsn} (杨煜波) \end{CJK}}
\affiliation{\hofstra}%
\affiliation{\ccq}

\author{Vitaly Gorelov}
\affiliation{\lsi}%
\affiliation{\etsf}%

\author{Miguel A. Morales}
\affiliation{\ccq}%

\author{Matteo Gatti}
\email{matteo.gatti@polytechnique.fr}%
\affiliation{\lsi}%
\affiliation{\etsf}%
\affiliation{\soleil}%

\author{Lucia Reining}
\email{lucia.reining@polytechnique.fr}%
\affiliation{\lsi}%
\affiliation{\etsf}%

\author{Shiwei Zhang}
\email{szhang@flatironinstitute.org}%
\affiliation{\ccq}%

\date{\today}

\begin{abstract}
    The local field factor $G(q)$ of the homogeneous electron gas describes how exchange and correlation modify the response to a static perturbation, and enters dielectric screening, effective electron--electron interactions, and the exchange-correlation kernel of time-dependent density functional theory. Accurate zero-temperature data for $G(q)$, however, remain limited to metallic densities, $r_s\leq10$. In this work, we present diffusion Monte Carlo calculations of the static density-density response function $\chi(q)$ and of $G(q)$ for $2\leq r_s\leq100$, and describe and discuss the methodological steps that enabled these calculations. We show how optimization of the perturbing potential in the trial wave function has an important effect, as a poor choice can reverse the sign of the response. A closed-form expression is provided for its optimal value, together with the density dependence of the linear-response window. We further show that backflow correlations leave $\chi(q)$ unchanged and that $G(q)$ depends only weakly on system size even at very low density, so that Slater--Jastrow wave functions and modest system sizes suffice. Together, these results provide a practical guide for future quantum Monte Carlo calculations of response properties of the electron gas and real materials. Our results agree with previous diffusion Monte Carlo data at metallic densities for $q>k_F$, and extend them below $k_F$. They also agree with diagrammatic quantum Monte Carlo at $r_s=2$. At lower densities, our results show that $G(q)$ develops a nonmonotonic structure at intermediate wave vectors, where the response approaches that of the noninteracting gas. Quantitative agreement is seen with finite-temperature path integral Monte Carlo data, suggesting that the previously reported finite-temperature structure is inherited from the ground state. Using a parametrization of $G(q)$ that reproduces our data throughout the liquid phase, we relate this structure to short-range order in the pair correlation function and the static structure factor.
\end{abstract}

\maketitle

\section{Introduction}
The response properties of the homogeneous electron gas (HEG) have been of considerable theoretical interest for over half a century~\cite{KUGLER1975,SINGWITOSI1981, GIULIANI2005,PINES1966}. They describe how the gas rearranges in response to an external potential of a given wavelength~\cite{KOHN1965a}. For a static external potential, this information is contained in the static linear density response function $\chi(q)$, and the local field factor $G(q)$ isolates the part of it due to exchange and correlation~\cite{GIULIANI2005}. Both enter directly into descriptions of dielectric screening~\cite{ICHIMARU1982}, effective electron--electron interactions~\cite{KUKKONEN1979,KUKKONEN2021}, and the exchange-correlation (xc) kernel of time-dependent density functional theory~\cite{PANHOLZER2018}. Early studies of the static response were based on analytical and diagrammatic schemes that satisfied the compressibility sum rule and the large-$q$ limit~\cite{HUBBARD1958,GELDART1966,SINGWI1968,GELDART1970,TOIGO1970,TOIGO1971,VASHISHTA1972,KUGLER1975,RAJAGOPAL1972,TRIPATHY1977,HOLAS1979,DEVREESE1980,BROSENS1980,UTSUMI1980,UTSUMI1980B,UTSUMI1981,ICHIMARU1982,HOLAS1987,ORTIZ1994,CHIHARA1979,MAHAN1981,GIULIANI2005}.

The first quantum Monte Carlo (QMC) calculations of the static response function $\chi(q)$ and the local field factor $G(q)$ of the gas at zero temperature were carried out in parallel by two groups in 1992. Sugiyama, Bowen, and Alder demonstrated the approach on the charged Bose gas~\cite{SUGIYAMA1992}, while Moroni, Ceperley, and Senatore applied it to the two-dimensional electron gas and liquid $^4$He~\cite{MORONI1992}. The first application to the three-dimensional electron gas was made by Bowen \textit{et al.}~\cite{BOWEN1994}, using variational Monte Carlo, fixed-node diffusion Monte Carlo (DMC), and released-node Green's function Monte Carlo to confirm that the fixed-node results are exact within statistical error bars. The definitive zero-temperature results were later provided by Moroni, Ceperley, and Senatore in their benchmark study~\cite{MORONI1995}, which yielded accurate DMC data for $\chi(q)$ and $G(q)$ across the metallic density range $r_s = 2$--$10$, alongside an analytical interpolation of $G(q)$ that satisfies the known asymptotic limits. Their data, calculated by adding an external perturbation to the Hamiltonian, showed that $G(q)$ varies smoothly between these limits with a crossover near $2k_F$. These calculations have served as the primary benchmark for describing exchange-correlation effects in materials for nearly three decades.

At finite temperature, there have been significant efforts especially using the path integral Monte Carlo (PIMC) approach. Dornheim \textit{et al.}\ have computed $\chi(q)$ and $G(q)$ for the strongly coupled, i.e., low-density, electron liquid at reduced temperatures $\Theta = T/T_F = 0.5$--$4$ for a wide range of $r_s$ values including $r_s = 20$, $r_s = 50$, $100$, and $200$ via imaginary-time correlation functions~\cite{DORNHEIM2024,DORNHEIM2020,DORNHEIM2019,SVENSSON2026}. Their findings show indications of structure in $G(q)$ at intermediate wave vectors; however, with thermal and correlation effects inextricably entangled at finite $T$, a clear origin of this feature could not be established.

Following the calculations of Moroni \textit{et al.}, several efforts to parametrize $G(q)$ have been made. Corradini \textit{et al.}\ fitted an analytical form to the Moroni \textit{et al.}\ data~\cite{CORRADINI1998}, which made $f_{\mathrm{xc}}(q)$ conveniently available for use in \textit{ab initio} calculations. This parametrization has been widely adopted ever since~\cite{OLEVANO1999,PALUMMO1999,CAZZANIGA2011,PANHOLZER2018,RAMAKRISHNA2021,DORNHEIM2019,AOUINA2025}. More recently, Kukkonen and Chen provided new results for $G(q)$ at $r_s=1$ and $r_s=2$ using diagrammatic QMC~\cite{KUKKONEN2021}. These results (along with the Moroni \textit{et al.}\ data) were then used by Kaplan and Kukkonen to construct an improved parametrization~\cite{KAPLAN2023}.

Crucially, despite these efforts on parametrizing $G(q)$, the underlying QMC data for the unpolarized gas do not extend beyond $r_s = 10$. Two distinct factors have obstructed the extension of the perturbative DMC approach of Moroni \textit{et al.} to lower densities. The first is cost. Each point of a $\chi(q)$ curve requires DMC calculations at several perturbation amplitudes, each preceded by an independent optimization of the trial wave function, which converges more slowly as the density decreases; a single curve at a single density can easily amount to $\sim\!10^3$ separate QMC calculations. This cost is met by modern implementations and computing resources~\cite{QMCPACK_1,QMCPACK_2}, which were instrumental for the present work. The second factor is not relieved by computing resources: several systematic errors that are controlled at metallic densities remain uncharacterized at $r_s\gtrsim10$. The usable range of perturbation strength to ensure linear response has not been established, which increases computational cost if one must repeat calculations at each density and several perturbation strengths to verify the validity of linear response. The effect of the nodal surface of the trial wave function on the extracted $\chi(q)$ has not been tested, and the finite-size effects in $\chi(q)$ have not been investigated at large $r_s$.

In this work, we present DMC calculations of the static response function $\chi(q)$ and the local field factor $G(q)$ of the three-dimensional electron gas at zero temperature, covering the entire liquid phase up to $r_s=100$. The physical implications of this new regime are discussed in a companion paper~\cite{GUNESprl}, where we show the emergence of a nonmonotonic structure in $G(q)$ at intermediate wave vectors and how it manifests as short-range order in the electron liquid, together with a new parametrization of $G(q)$ that can be used for many-body calculations. In this paper, we focus on the methodology that made those findings possible, addressing in turn each of the obstacles above. We determine the optimal variational parameters of the trial wave function and reduce them to a closed form valid across the whole density range, so that the optimization does not have to be repeated for every calculation; we establish how far the perturbation amplitude can be increased while remaining within the linear response regime as the density decreases; we examine whether more sophisticated forms of the trial wave function meaningfully alter the extracted susceptibility, and we investigate the finite-size effects at low density.

The rest of the paper is organized as follows. Section~\ref{sec:background} gives a complete description of the methodology, including how the response function is extracted, how the regime of linear response is determined, how the trial wave function is constructed and optimized, how finite-size and time-step errors are treated, and how $G(q)$ is parametrized. Section~\ref{sec:results} presents the results for $\chi(q)$ and $G(q)$ for $r_s=2$--$100$ and highlights the features that emerge at large $r_s$. Subsection~\ref{sec:discussion} relates these features to the structure of the strongly coupled electron liquid, as seen in its pair correlation function and static structure factor, and to its evolution toward the Wigner crystal. We conclude in Sec.~\ref{sec:conclusion}.

\section{Method}\label{sec:background}

\subsection{Static response from a cosine perturbation}\label{sec:response}

We extract the response properties by subjecting an otherwise homogeneous electron gas to a weak static cosine perturbation,
\begin{equation}\label{eqn:v_ext}
    v_{\mathrm{ext}}(\mathbf r)=2v_\mathbf{q}\cos(\mathbf q\!\cdot\!\mathbf r),
\end{equation}
where $v_\mathbf{q}$ is the amplitude of the perturbation. The response function $\chi(q)$ can then be extracted by expanding the perturbed energy (per particle) in powers of $v_\mathbf{q}$. Within linear response, the ground-state energy is
\begin{equation}\label{eqn:expansion}
    E(v_\mathbf{q}) = E_0+\frac{\chi(q)}{n_0}v^2_\mathbf{q} + \cdots,
\end{equation}
where $E_0$ is the unperturbed ground-state energy per particle (i.e., $v_\mathbf{q}=0$), and $\chi(q)$ is the static density-density response function. Higher-order terms carry information about the higher-order response, which we do not explore in this work. By fitting the energy to this form, one can extract $\chi(q)$ from the coefficient of the quadratic term.

A crucial quantity that can be obtained directly from $\chi(q)$ is the local field factor $G(q)$, which encodes the exchange and correlation effects beyond the random-phase approximation (RPA). It is defined via the Dyson equation,

\begin{equation}\label{eq:G}
    G(q) = -\frac{f_{\mathrm{xc}}(q)}{v_c(q)}  = 1 + \frac{\chi^{-1}(q) -  \chi^{-1}_0(q)}{v_c(q)},
\end{equation}
where $f_{\mathrm{xc}}(q)$ is the exchange-correlation kernel, $v_c(q) = 4\pi/q^2$ is the Coulomb interaction, and $\chi_0(q)$ is the static response function of the noninteracting system~\cite{GIULIANI2005}. The noninteracting and RPA limits are reached by setting $G(q)=1$ and $G(q)=0$, respectively. Since $\chi(q)$ will be obtained from stochastic estimates of the energy $E(v_\mathbf{q})$, it carries a statistical error, which propagates into $G(q)$ with a prefactor proportional to $\left(q/\chi(q)\right)^2$ that inevitably diverges in both the low- and high-$q$ limits. However, this is of no practical concern since $G(q)$ is known analytically in both limits~\cite{HOLAS1987}. The low-$q$ limit of the local field factor is given by $\lim_{q \to 0} G(q) = A(q/k_F)^2$ whereas the high-$q$ limit is given by $\lim_{q \to \infty} G(q) = C(q/k_F)^2 + B$. All quantities $A,B, C$ are known expressions and parametrizations as a function of $r_s$~\cite{MORONI1995,CORRADINI1998,ORTIZ1994,HOLAS1987}. We provide the explicit expressions for $A(r_s)$, $B(r_s)$, and $C(r_s)$ in Appendix~\ref{app:limits}.
\begin{figure}[b]
    \centering
    \includegraphics[width=\linewidth]{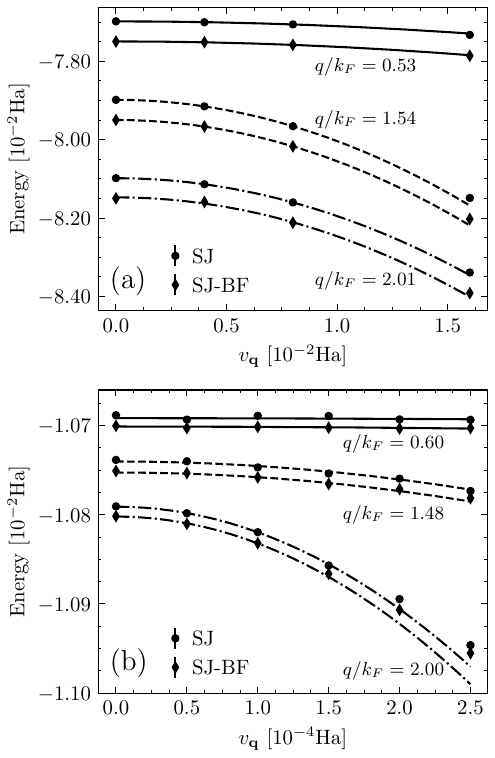}
    \caption{Comparison of $E(v_\mathbf{q})$ obtained using DMC with and without backflow transformation for (a) $r_s=5$, $N_e=162$, $q/k_F=0.53$ (solid), $1.54$ (dashed), $2.01$ (dash-dotted) and (b) $r_s=70$, $N_e=38$, $q/k_F=0.60$ (solid), $1.48$ (dashed), $2.00$ (dash-dotted). For better visibility, the data at the second and third $q$ are shifted downward by $0.002$ and $0.004$~Ha in (a), and by $5\times10^{-5}$ and $10^{-4}$~Ha in (b). Lines represent a quadratic fit. Error bars are smaller than the symbols.}
    \label{fig:backflow}
\end{figure}

To compute the perturbed energies $E(v_\mathbf{q})$, we employ the state-of-the-art fixed-node DMC method~\cite{CEPERLEY1978,CEPERLEY1980,MORONI1995}, implemented in \textsc{QMCPACK}~\cite{QMCPACK_1,QMCPACK_2}. DMC is the gold standard for computing ground-state properties of quantum many-body systems, and it has been extensively used as a benchmark for the homogeneous electron gas~\cite{CEPERLEY1980,TANATAR1989}. Modifications were made to include the cosine perturbation to the Hamiltonian. Once $E(v_\mathbf{q})$ is obtained, $\chi(q)$ is extracted from the energy by performing a least-squares fitting procedure to the form of Eq.~\eqref{eqn:expansion}. The procedure is summarized in Fig.~\ref{fig:backflow} for two different densities, $r_s=5$ and $r_s=70$, and for three different wave vectors. We carefully select the range of $v_\mathbf{q}$ used to fit $E(v_\mathbf{q})$ and extract $\chi(q)$ so that it remains in the linear-response regime, where the signal-to-noise ratio is high. The resulting maximum perturbation is parametrized by the fit $v_\mathbf{q}^\mathrm{max} = 0.1212\,r_s^{-1.43}$~Ha, where the exponent is determined by the noninteracting problem and the prefactor is fitted to QMC data. More details of this selection process can be found in Appendix~\ref{app:vq_range}.

\subsection{Trial wave function}
We construct our trial wave function using both Slater--Jastrow (SJ) and Slater--Jastrow-backflow (SJ-BF) forms~\cite{JASTROW1955,LEE1981,SCHMIDT1981,KWON1993,KWON1998}. The former is the standard form used in many QMC calculations in which the nodal surface is fixed by the Slater determinant of single-particle orbitals, while the latter allows for modifications of the single-particle coordinates in the Slater determinant by introducing a displacement for the $j$th electron, which depends on the positions of all other electrons. Such a transformation introduces additional variational parameters that can be optimized to further minimize the energy. To compare both forms and verify that our choice of trial wave function does not bias the extraction of $\chi(q)$, we performed a subset of calculations with and without a backflow transformation. We find that while the backflow transformation significantly improves the energies obtained from DMC, it does not affect the extraction of $\chi(q)$, which is related to the curvature of the energy with respect to the perturbation amplitude $v_\mathbf{q}$, and not the absolute value of the energy itself. Figure~\ref{fig:backflow} demonstrates this point for (a) $r_s=5$ with $N_e=162$, and for (b) $r_s=70$ with $N_e=38$, at three wave vectors. The resulting $\chi(q)$ values obtained with and without backflow transformation are in agreement within error bars, even though the energies themselves are significantly different. Thus, throughout this work, we will mainly use the Slater--Jastrow trial wave function without backflow transformation (unless stated otherwise) to extract $\chi(q)$, which is computationally more efficient and shown not to bias our results. The SJ trial wave function is constructed as $\Psi_T(\mathbf{R}) = e^{J(\mathbf{R})} \det[\phi_i(\mathbf{r}_j)]$, where $J(\mathbf{R})$ is the Jastrow factor that captures electron-electron correlations, and $\det[\phi_i(\mathbf{r}_j)]$ is the Slater determinant of noninteracting single-particle wave functions. For the latter, we will use the one of noninteracting single-electrons~\cite{q-e_short} perturbed by an effective external potential of the same form, %
\begin{equation}\label{eq:alpha}
    v^\mathrm{eff}_{\mathrm{ext}}(\mathbf{r})=\alpha v_{\mathrm{ext}}(\mathbf{r}).
\end{equation}
\begin{figure}[t]
    \centering
    \includegraphics[width=\linewidth]{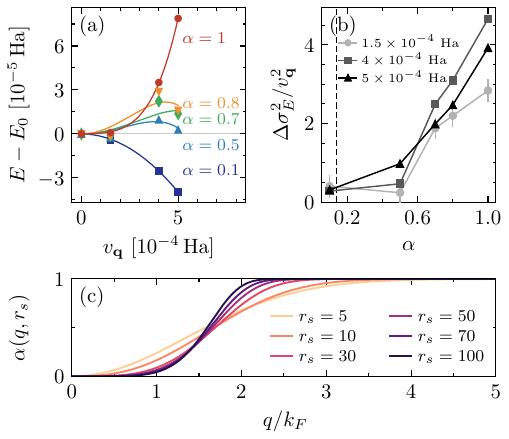}
    \caption{(a) Energy per electron $E(v_\mathbf{q})$ relative to the ground state energy $E_0$ as a function of the perturbation amplitude $v_\mathbf{q}$ for trial wave functions with different $\alpha$, at $r_s=50$, $N_e=54$, and $q=1.20\,k_F$. Solid curves are quartic fits. (b) Excess variance, $\Delta \sigma_E^2 = \sigma_E^2(v_\mathbf{q}) - \sigma_E^2(0)$, of the local energy per electron normalized by $v_\mathbf{q}^2$ versus $\alpha$ for different amplitudes. Vertical dashed line indicates the optimal value of $\alpha$ given by Eq.~\eqref{eq:opt_alpha}. Error bars are reported on the symbols. (c) Optimized $\alpha$ as a function of $q/k_F$ for different values of $r_s$. }
    \label{fig:alpha_scan}
\end{figure}
By construction, $v_{\mathrm{ext}}^\mathrm{eff}$ does not have to be equal to the amplitude of the external potential $v_\mathbf{q}$, so $\alpha$ must be treated as a variational parameter that yields the minimum energy. Therefore, the optimization is to be achieved in two steps. First, one performs variational Monte Carlo (VMC) calculations to minimize the energy computed from the Hamiltonian perturbed by Eq.~\eqref{eqn:v_ext}, with respect to the parameters in the Jastrow factor, for a given $\alpha$ in Eq.~\eqref{eq:alpha}. This procedure is then repeated for different $\alpha$ values to obtain the VMC energy as a function of $\alpha$, i.e., $E^v(\alpha)$, and extract the optimal $\alpha$. In this work, we find that the $\alpha$ that minimizes the variational energy is different for each value of $\mathbf{q}$ and $r_s$ and needs to be optimized for each case separately. We emphasize that the $\alpha$-optimization is crucial to obtain reliable results for $\chi(q)$ for all $r_s$ values. This is demonstrated in Fig.~\ref{fig:alpha_scan}(a) for $r_s=50$ and $q=1.2k_F$. A poorly chosen $\alpha$ changes not only the energy but also the curvature of $E(v_\mathbf{q})$, whose sign can even be reversed, and gives rise to an unphysical $\chi(q)>0$. As $\alpha$ is reduced (for this particular wave vector and density), the curvature of $E(v_\mathbf{q})$ converges to a stable value, and yields the correct $\chi(q)$ at $\alpha \simeq 0.15$. Figure~\ref{fig:alpha_scan}(b) shows that the variance of the local energy decreases accordingly as $\alpha$ approaches the optimal value, i.e., the minimum of $E^v(\alpha)$ (vertical dashed line). Performing over $700$ VMC calculations using the above approach, we found that the optimal $\alpha$ is reproduced by the following form for $r_s=2$--$100$:
\begin{equation}\label{eq:opt_alpha}
    \alpha(q,r_s) = 1 - e^{-\alpha_0(r_s)\,(q/k_F)^{\alpha_1(r_s)}},
\end{equation}
where $\alpha_0(r_s) = 0.7276\,r_s^{-0.6050}$ and $\alpha_1(r_s) = 1.1603\,r_s^{0.3513}$. Figure~\ref{fig:alpha_scan}(c) shows $\alpha(q,r_s)$ as a function of $q/k_F$ for six different values of $r_s$. Once the optimal $\alpha$ is obtained, we perform DMC calculations to obtain the perturbed energies $E(v_\mathbf{q})$, which are then used to extract $\chi(q)$.

\subsection{Size effects and time-step error}\label{subsec:fse}
It has been shown that the local field factor $G(q)$ is much less sensitive to finite-size effects than the energy~\cite{MORONI1995} for $r_s\leq10$. This is due to the fact that exchange and correlation effects are short-ranged. Therefore, one can assume $G(\mathbf{q}, N) \simeq G(q,\infty)$ to mitigate finite-size effects in our calculations as follows

\begin{equation}\label{eq:fse}
    \chi^{-1}(\mathbf{q},\infty)=\chi^{-1}(\mathbf{q},N) + \Delta\chi^{-1}_0(\mathbf{q}),
\end{equation}
where $\Delta\chi_0^{-1}(\mathbf{q}) = \chi_0^{-1}(\mathbf{q},\infty) -\chi_0^{-1}(\mathbf{q},N)$ represents the finite-size correction in the noninteracting response function. Throughout this work, we use Eq.~\eqref{eq:fse} to correct the finite-size effects in $\chi(q)$. At low-$r_s$ values, Eq.~\eqref{eq:fse} is known to hold quite well~\cite{MORONI1995}. However, at high-$r_s$ values, such a claim is questionable and merits further investigation.

\begin{figure}[b]
    \centering
    \includegraphics[width=\columnwidth]{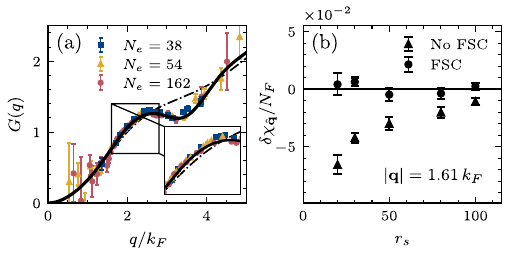}
    \caption{(a) The local field factor $G(q)$ at $r_s=50$ for different system sizes. The inset shows the zoomed-in view around $q=2k_F$. (b) The difference between two $\chi(q)$-values that have the same $|\mathbf{q}|$ but different directions, i.e., $ \chi\left(\mathbf{q} = (2,2,1)\frac{2\pi}{L}\right) - \chi\left(\mathbf{q} = (3,0,0)\frac{2\pi}{L}\right)$ as a function of $r_s$ for $N_e=54$. The triangles and circles correspond to the raw (No FSC) and finite-size-corrected (FSC) $\chi(q)$ values, respectively.}
    \label{fig:fse}
\end{figure}

In this subsection, we investigate the validity of this hypothesis at large $r_s$ by testing two finite-size features of the Fermi sphere: its resolution, set by $N_e$, and its isotropy at fixed $N_e$. Figure~\ref{fig:fse}(a) shows the local field factor $G(q)$ at $r_s=50$ for different system sizes: $N_e=38$ (squares), $N_e=54$ (triangles), and $N_e=162$ (circles). The inset zooms into the region around $2k_F$. Overall, the DMC data points for different system sizes collapse onto a single curve, confirming the insensitivity of $G(q)$ to finite-size effects, even at large $r_s$ values. Although the statistical uncertainty of $G(q)$ at $q > 4k_F$ and $q \lesssim 1.6\,k_F$ is too large to resolve differences between system sizes, both the small- and large-$q$ limits are analytically known (given in Appendix~\ref{app:limits}). Figure~\ref{fig:fse}(b) shows the difference between two $\chi(q)$-values that have the same $|\mathbf{q}|$ but different directions, i.e., $ \delta\chi_{\hat{\mathbf{q}}} = \chi\left(\mathbf{q} = (2,2,1)\frac{2\pi}{L}\right) - \chi\left(\mathbf{q} = (3,0,0)\frac{2\pi}{L}\right)$ as a function of $r_s$. Since in the thermodynamic limit, $\delta\chi_{\hat{\mathbf{q}}}$ should vanish, any nonzero value of $\delta\chi_{\hat{\mathbf{q}}}$ indicates the presence of finite-size effects. This is shown by the triangles in Fig.~\ref{fig:fse}(b), which correspond to the raw $\chi(q)$ values. If Eq.~\eqref{eq:fse} captures the dominant finite-size error, the corrected $\chi(q)$ should give $\delta\chi_{\hat{\mathbf q}}=0$. This is shown by the circles in Fig.~\ref{fig:fse}(b), where the difference is effectively reduced to zero within error bars. The anisotropy of $\chi(\mathbf q,N)$ at fixed $N_e$ is therefore entirely due to $\chi_0(\mathbf q,N)$, and it is removed by the correction in Eq.~\eqref{eq:fse}.%

\begin{figure}[b]
    \centering
    \includegraphics[width=\columnwidth]{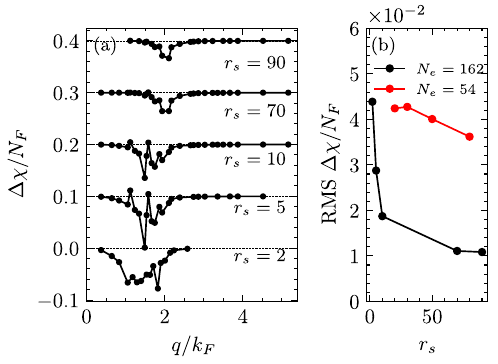}
    \caption{Finite-size correction to the static response, $\Delta\chi \equiv \chi(q,\infty) - \chi(\mathbf{q},N)$, obtained from Eq.~\eqref{eq:fse} and given in units of $N_F$. (a) $\Delta\chi/N_F$ as a function of $q/k_F$ for $N_e=162$ at $r_s = 2$--$90$; curves are offset vertically by $0.1$, with dashed lines marking zero for each $r_s$. (b) Root mean square of $\Delta\chi/N_F$ over all computed $q$ points as a function of $r_s$, for $N_e=162$ (black) and $N_e=54$ (red).}
    \label{fig:chi_fse}
\end{figure}

The remaining question is how large this correction is, and how it changes with increasing coupling. For this, we show in Fig.~\ref{fig:chi_fse}(a) the finite-size correction to the static response, $\Delta\chi \equiv \chi(q,\infty) - \chi(\mathbf{q},N)$, obtained from Eq.~\eqref{eq:fse} and given in units of $N_F=k_F/\pi^2$. For all densities, $\Delta\chi$ is localized in $q$ and vanishes at small and large $q$. The correction is largest at high density around $q\simeq1.5k_F$, and it becomes smaller at low density, where it shifts to $q\simeq2k_F$. It is predominantly negative, i.e., the finite system underestimates $|\chi(q)|$ where the correction is largest, and its magnitude always remains below $0.1N_F$ (where $N_F=-\chi_0(q\to0)$ sets the natural scale of the response). Its root-mean-square [Fig.~\ref{fig:chi_fse}(b)] over all computed wave vectors decreases from $\approx4\times10^{-2}$ at $r_s=2$ to $\approx1\times10^{-2}$ at $r_s\gtrsim70$ for $N_e=162$, and approaches the typical statistical uncertainty of $\chi(q)$ (around $0.5\times10^{-2}$ at $r_s=90$). $N_e=54$ shows the same decreasing trend. The finite-size correction to $\chi(q)$ from Eq.~\eqref{eq:fse} is therefore modest at all densities, and it decreases further as the coupling increases. Twist averaging is an alternative route to removing one-body finite-size errors, but is more expensive, since it requires an independent calculation for each twist. In Appendix~\ref{app:twist}, we show that at fixed $N_e$ (canonical twist averaging), it approaches our corrected results to within 2--3\%, with a small residual bias due to keeping the number of electrons fixed.

\begin{figure}[h]
    \centering
    \includegraphics[width=\linewidth]{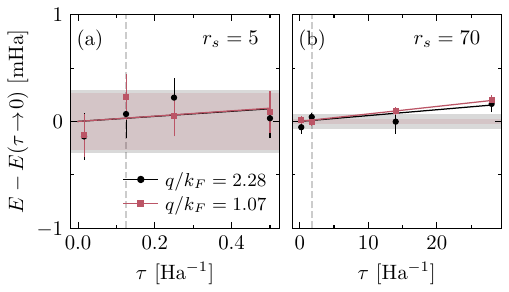}
    \caption{Time-step bias $E(\tau)-E(\tau\to0)$ in the energy of the perturbed gas at $N_e=54$, for two wave vectors. (a) $r_s=5$ perturbed with $v_\mathbf{q} = 4$ mHa, (b) $r_s=70$ perturbed with $v_\mathbf{q} = 0.1$ mHa. Red and black correspond to $q/k_F=1.07$ and $q/k_F=2.28$, respectively. Lines are weighted linear fits, and the shaded bands indicate the error bar of $E(\tau\to0)$. The vertical dashed lines indicate the time step used in the production runs.}
    \label{fig:time_step_extrapolation}
\end{figure}

Besides finite-size effects, the DMC energies carry a time-step error. Since $\chi(q)$ is extracted from small energy differences between perturbed states, any time-step bias that depends on $q$ or $v_\mathbf{q}$ would directly bias the curvature of $E(v_\mathbf{q})$. Figure~\ref{fig:time_step_extrapolation} therefore shows the time-step extrapolation for the perturbed gas at two wave vectors and two densities. The DMC energies exhibit a linear dependence on the time step $\tau$ as it approaches zero. For $r_s=5$, the energy is converged within error bars for $\tau \leq 0.5$ Ha$^{-1}$, while for $r_s=70$, the energy is converged for $\tau \leq 14$ Ha$^{-1}$. This is consistent with the fact that for such a large $r_s$ value, the system is in a strongly correlated regime where the energy scales shrink, and the DMC algorithm can take larger time steps without losing accuracy. In our production runs, we have used $\tau=r_s/40$ Ha$^{-1}$ (vertical dashed lines), which is well within the converged regime for all $r_s$ values considered in this work.

\subsection{Analytical fit}\label{sec:fit}
\begin{figure*}[ht]
    \centering
    \includegraphics[width=\linewidth]{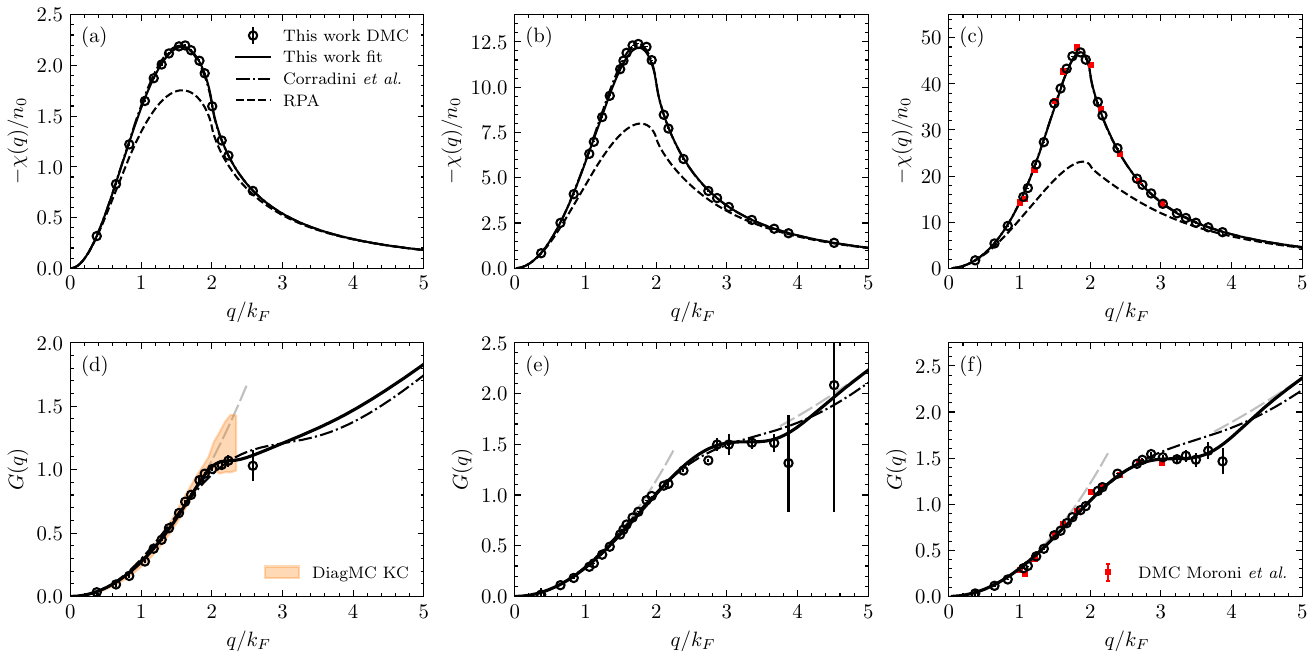}
    \caption{Static response function $\chi(q)$ (top) and local field factor $G(q)$ (bottom) for $N_e=162$ and $r_s=2,5,10$, from left to right. This work: circle data points and black solid line; orange shaded region: Ref.~\cite{KUKKONEN2021}; red squares: Ref.~\cite{MORONI1995}; dash-dotted line: Ref.~\cite{CORRADINI1998}; dashed line (top): RPA; gray dashed lines (bottom): low- and high-$q$ limits of $G(q)$.}
    \label{fig:G_chi_rs2_5_10}
\end{figure*}

All solid curves labeled ``This work'' that we will present in Sec.~\ref{sec:results} are obtained from a single parametrization of $G(q)$, introduced in the companion paper~\cite{GUNESprl}. It interpolates between a low- and a high-$r_s$ form,
\begin{equation}\label{eqn:G_fit}
    G(q) = (1-s)\,G_\mathrm{lo}(q) + s\,G_\mathrm{hi}(q), \quad s=\tanh\!\left(r_s^4/250\right),
\end{equation}
where both terms take the form of Kaplan and Kukkonen~\cite{KAPLAN2023},
\begin{equation}\label{eqn:G_KK}
    G_\mathrm{lo,hi}(q) = \underbrace{A x^{2} h}_{\text{small }x}
    + \underbrace{\theta x^{n+2} h}_{\text{interm. }x}
    + \underbrace{(Cx^{2}+B)(1-h)}_{\text{large }x},
\end{equation}
with $x=q/k_F$, the exact coefficients $A$, $B$, and $C$ given in Appendix~\ref{app:limits}, and the crossover function
\begin{equation}\label{eqn:h}
    h = \frac{e^{a_3a_4}-1}{e^{a_3x^4/16}+e^{a_3a_4}-2}.
\end{equation}
The function $h$ decays from $1$ to $0$, with position set by $a_4$ and width by $a_3$. Since $h\to1$ for $x\to0$ and the intermediate term vanishes faster than $x^2$ for $n>0$, Eq.~\eqref{eqn:G_KK} satisfies the compressibility sum rule and the large-$q$ limit exactly, at every $r_s$. The low-$r_s$ form $G_\mathrm{lo}(q)$ is the original parametrization of Ref.~\cite{KAPLAN2023}, with $n=4$ and $\theta=a_0+a_1e^{-a_2r_s}$. Kaplan and Kukkonen reported $a_0$ and $a_2$ with sizable uncertainties, due to the growing error bars of the diagrammatic QMC data above $2k_F$ [see orange shaded area in Fig.~\ref{fig:G_chi_rs2_5_10}(d)]. Refitting these two parameters to include our DMC data, we obtain $a_0=-0.0046277$, within the original error bar, and $a_2=0.622624$, at the edge of the original uncertainty, and keep the remaining parameters at their published values, $a_1=0.0155766$, $a_3=3.516054$, and $a_4=1.015830$.

The high-$r_s$ form $G_\mathrm{hi}(q)$ keeps the same structure, but lets the exponent evolve with density, $n=a'_5+a'_6e^{-a'_7r_s}$, while $\theta=a'_0+a'_1e^{-a'_2r_s}$ keeps the original form with refitted parameters. The crossover $h$ is moved to larger $x$ through $a'_4$, with a broader width $a'_3$. At large $r_s$, $\theta$ becomes negative, and the intermediate term allows for a possible nonmonotonic behavior of $G(q)$, which we will discuss in Sec.~\ref{sec:results}. The values of the primed parameters are given in Ref.~\cite{GUNESprl}. The switching function $s$ is negligible at $r_s=1$, reaches $s\simeq0.06$ at $r_s=2$, and is close to unity for $r_s\geq5$, so that $G_\mathrm{lo}(q)$ is recovered in the regime of metallic densities, and $G_\mathrm{hi}(q)$ describes the strongly coupled regime.

\section{Results}\label{sec:results}
We have performed DMC calculations for the unpolarized three-dimensional electron gas between $2\leq r_s \leq 100$, with $N_e=38$, 54, and 162 electrons. The values of $r_s$ are chosen to cover the entire liquid phase of the electron gas, and perhaps even beyond the Wigner crystallization transition, since current estimates for the transition vary between $r_s \simeq 87$ and $r_s \simeq 113$~\cite{AZADI2022,CEPERLEY1980, DRUMMOND2004,HOLZMANN2020}. Note, however, that our ansatz is limited to represent the liquid phase only, even if it is metastable. Our results are compared to the existing Corradini \textit{et al.} parametrization, which becomes an extrapolation at $r_s>10$, in order to highlight new features emerging in the strongly coupled regime.

\begin{figure*}[ht]
    \centering
    \includegraphics[width=\linewidth]{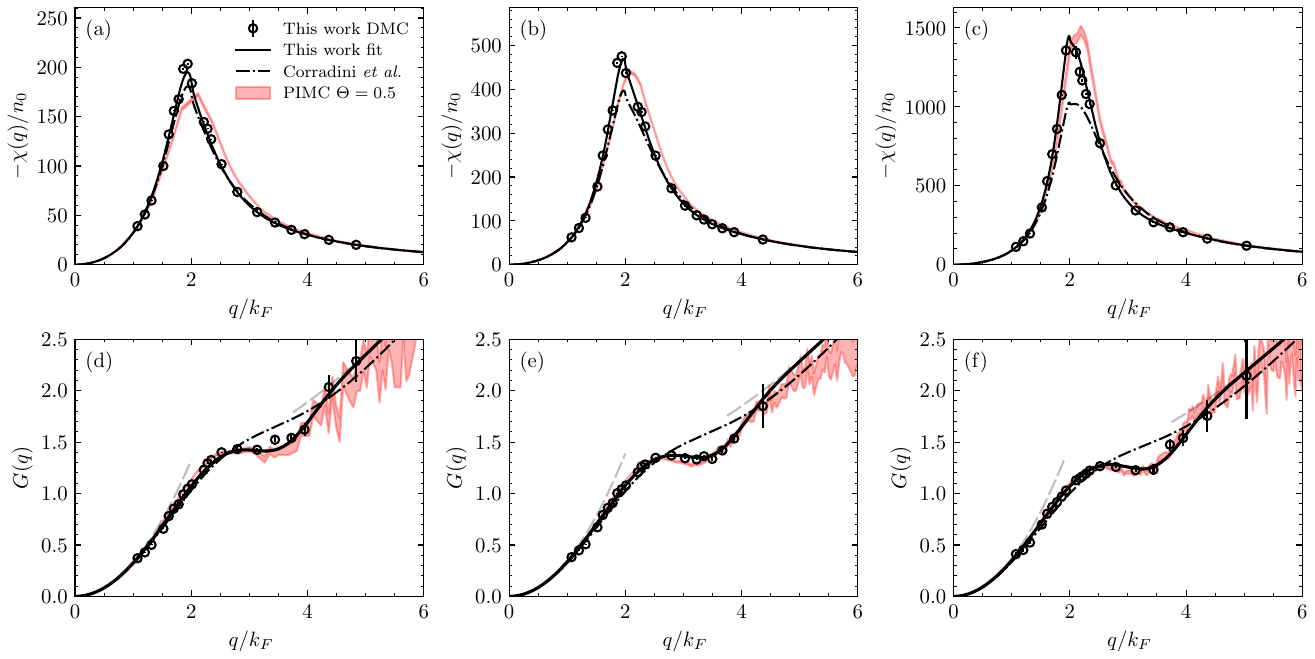}
    \caption{Static response function $\chi(q)$ (top) and local field factor $G(q)$ (bottom) for $r_s=20,30,50$, from left to right, using $N_e=54$ throughout, with additional $N_e=162$ points at $r_s=30$ and $50$. This work: circle data points and black solid line; dash-dotted line: Ref.~\cite{CORRADINI1998} (extrapolated); red shaded: (lowest) finite-$T$ PIMC from Ref.~\cite{DORNHEIM2020}. Gray dashed lines represent the low- and high-$q$ limits of $G(q)$.}
    \label{fig:G_chi_rs20_30_50}
\end{figure*}

\subsection{Static response function and local field factor from DMC}\label{sec:highdensity}
We present our results for the static density response function $\chi(q)$ and the local field factor $G(q)$, using Eq.~\eqref{eq:G}, for $r_s=2,5,10$ in the upper and lower panels of Fig.~\ref{fig:G_chi_rs2_5_10}, respectively. The DMC data are shown as circles, while the dash-dotted curve is the analytical parametrization of Ref.~\cite{CORRADINI1998} designed specifically for these $r_s$ values to fit the data from Ref.~\cite{MORONI1995}. As a first demonstration, our results for $\chi(q)$ are in excellent agreement with the existing parametrizations fitted at these same densities. The response of the system vanishes as $q \to 0$ quadratically (governed by the compressibility limit) due to the long-range nature of the Coulomb interaction, and peaks at $q \simeq 2k_F$ before approaching the RPA limit at large $q$. The peak becomes more pronounced with decreasing density, while its departure from RPA is an indication of growing exchange-correlation effects. Importantly, our simulations provide the first DMC data for $q<k_F$, and they are consistent with the compressibility limit. This is important for numerous applications~\cite{KUKKONEN2021,KAPLAN2023,NEPAL2020,PELLEGRINI2023,LI2026}.

On the other hand, the case of $G(q)$ is more delicate, as it isolates the exchange-correlation contribution that may not be immediately apparent in $\chi(q)$. We show $G(q)$ in Figs.~\ref{fig:G_chi_rs2_5_10}(d)--(f), where the small- and large-$q$ limits are given as gray dashed lines. For $r_s=2$, we present additional data (orange shaded region) obtained from diagrammatic QMC~\cite{KUKKONEN2021} which show a hump in $G(q)$ near $2k_F$ that is not present in previous parametrizations. This feature is included in the parametrization of Kaplan and Kukkonen~\cite{KAPLAN2023}, i.e., the intermediate term in Eq.~\eqref{eqn:G_KK}. However, the amplitude of the hump is uncertain due to increasingly large error bars in the QMC data for $q>2k_F$. Our DMC data confirm the presence of such a feature, but we find the enhancement to be quite small. Moreover, our significantly smaller error bars (obtained using a backflow transformation for $q>1.75k_F$ and $r_s=2$ to reduce the variance) resolve the discrepancy between DMC and diagrammatic QMC reported in Ref.~\cite{KUKKONEN2021}, as both methods now agree within error bars. Our DMC results, combined with the diagrammatic QMC data, therefore allow us to determine the parameters of Kaplan and Kukkonen more precisely (Sec.~\ref{sec:fit}). %

With the smaller error bars, the large-wave-vector regime becomes reliably accessible. Starting from $r_s=5$, we find that $G(q)$ exhibits a softening above $2k_F$ and deviates from the monotonic behavior fitted by the existing parametrizations, before approaching the high-$q$ limit. This is because the intermediate-$q$ regime was essentially unexplored by previous DMC data (e.g., at the lowest density, $r_s=10$, the data of Moroni \textit{et al.}, shown as red squares in Fig.~\ref{fig:G_chi_rs2_5_10}(f), do not extend beyond $q \simeq 3k_F$, but the feature is discernible in their data as well), so the parametrizations of Moroni \textit{et al.} and Corradini \textit{et al.} were designed as smooth crossovers between the known low- and high-$q$ limits~\footnote{In fact, the same behavior is also reported in Ref.~\cite{CORRADINI1998}, where the authors report the unpublished $r_s=100$ data of Moroni \textit{et al.} for a fully polarized gas.}. Our parametrization, designed to capture the nonmonotonic behavior of $G(q)$ in the intermediate-$q$ region, accurately reproduces the DMC data throughout the whole range.

This feature is a precursor of  a more pronounced nonmonotonicity due to growing exchange-correlation effects, as we show in Fig.~\ref{fig:G_chi_rs20_30_50} at $r_s=20,30,50$. At stronger coupling, we see that $G(q)$ develops a clear structure that gives rise to a local minimum. In contrast to the existing parametrization proposed by Corradini \textit{et al.}  (dash-dotted line, which is an extrapolation in this density range) that assumed monotonic behavior, the amplitude of this structure grows systematically with $r_s$ and $G(q)$ approaches $1$ at its local minimum. The position of the minimum shifts only slightly with density, moving from $q \simeq 3.3k_F$ to $3.2k_F$ as $r_s$ increases. The suppression toward $G(q) \simeq 1$ implies that the response function $\chi(q)$ tends toward the noninteracting response around the same $q$-value. This is an interesting feature, since it implies that the effective interaction $v_c(q)[1-G(q)]$ between electrons nearly vanishes (see Eq.~\eqref{eq:G}). The exchange-correlation effects cancel the Coulomb interaction, and the gas responds to a perturbation of this wave vector like a noninteracting system despite being strongly correlated. Our parametrization (Sec.~\ref{sec:fit}) is constructed to capture this nonmonotonic behavior. Similar observations in the Bose liquid, the one-component plasma, and the 2D electron gas have been reported in the literature, with various possibilities attributed to strong liquid~\cite{MITHEN2012} or Coulomb interaction~\cite{SENATORE1999}, scattering of electron pairs of opposite spin~\cite{WARDANA2003}, the $2k_F$-singularity in the response function responsible for Friedel oscillations~\cite{CORRADINI1998}, etc. In our companion paper~\cite{GUNESprl}, as well as in Sec.~\ref{sec:discussion}, we show how this feature in $G(q)$ relates to short-range order in the electron liquid as it approaches the Wigner crystallization transition.

Beyond the local minimum, we see in Fig.~\ref{fig:G_chi_rs20_30_50} that $G(q)$ approaches its large-$q$ limit. Whether it does so with further, smaller oscillations cannot be resolved with our current data. Similar nonmonotonic structure in $G(q)$ has been reported in PIMC calculations~\cite{DORNHEIM2020,DORNHEIM2024}, but whether it is a ground-state property or a finite-temperature effect has remained an open question, since the fermion sign problem prevents PIMC from reaching low temperatures~\cite{DORNHEIM2018,DORNHEIM2019}. Therefore, for the high-$r_s$ panels of Fig.~\ref{fig:G_chi_rs20_30_50}, we also show the finite-temperature PIMC results of Ref.~\cite{DORNHEIM2020} for $\Theta = 0.5$, which allow a direct comparison between ground-state and finite-temperature results. As panels (d)--(f) show, the PIMC data are in quantitative agreement with our $T = 0$ results, demonstrating that this structure is inherited from the ground state of the strongly coupled electron liquid rather than being a thermal effect.\footnote{At $r_s = 100$ (not shown), close to Wigner crystallization, the PIMC data for $G(q)$ no longer agree quantitatively, although the same structure is present.}. Moreover, via Eq.~\eqref{eq:G}, the agreement in $G(q)$ implies that the thermal changes in $\chi^{-1}(q)$ and $\chi_0^{-1}(q)$ largely cancel, i.e., that the temperature dependence of $\chi(q)$ at $\Theta = 0.5$ is carried predominantly by $\chi_0(q)$.
This is consistent with panels (a)--(c) of Fig.~\ref{fig:G_chi_rs20_30_50}, where $\chi(q)$ itself, in which no such cancellation occurs, agrees less well between PIMC and DMC. This suggests that $G(q)$ determined at moderate temperatures can be used to obtain $\chi(q)$ at $T\to0$, and vice versa.

\begin{figure}[t]
    \centering
    \includegraphics[width=\linewidth]{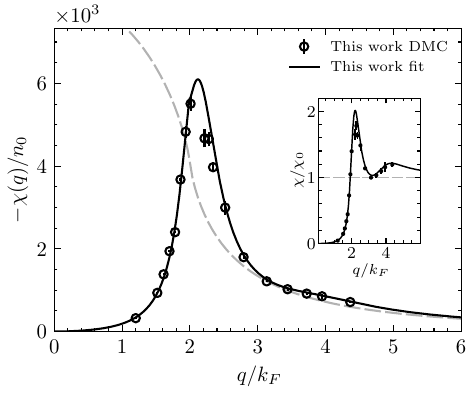}
    \caption{Static response function $\chi(q)$ at $r_s=100$ from DMC calculations (circles) with $N_e=54$. The solid line is obtained from the parametrization of $G(q)$ in Eq.~\eqref{eqn:G_fit}, and the dashed gray line is the noninteracting response $\chi_0(q)$. The inset shows the ratio $\chi(q)/\chi_0(q)$.}
    \label{fig:rs100}
\end{figure}

Finally, we show in Fig.~\ref{fig:rs100} the results for $r_s=100$, the lowest density we consider. The inset shows the ratio $\chi(q)/\chi_0(q)$. At such low density, whether the ground state of the electron gas is liquid or crystalline is still not precisely known, and our data show interesting features that could be related to the onset of such a transition. First, we see that the suppression of $\chi(q)$ toward the noninteracting response (dashed gray line) is the strongest among all densities considered, with $\chi(q_\mathrm{min})/\chi_0(q_\mathrm{min})=1.00\pm0.02$ at $q\simeq3.2k_F$ (see inset). Second, we also see a secondary peak in $\chi/\chi_0$ at about twice the wave vector of the first peak, which leaves a direct imprint on the excitation spectrum of the electron liquid~\cite{GUNESprl}.

\subsection{Relation to short-range order}\label{sec:discussion}
The structure in $G(q)$ emerges in the same range of $r_s$ in which $g(r)$ and $S(q)$ develop pronounced correlation features, suggesting that the two are related. Here, we show that the feature in $G(q)$ is a manifestation of the same physical phenomenon that shapes $g(r)$ and $S(q)$, commonly referred to as short-range order.

\subsubsection*{Electron liquid with short-range order}
\begin{figure}[t]
    \centering
    \includegraphics[width=\linewidth]{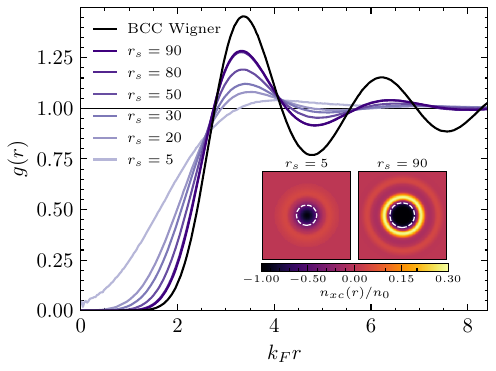}
    \caption{Short-range versus long-range order in the electron gas, seen through the radially averaged pair correlation function $g(r)$. All liquid-phase curves are from independent DMC calculations of the unperturbed gas with $N_e=162$, and the Wigner crystal (long-range order) at $r_s=110$ is from Ref.~\cite{DRUMMOND2004}. The insets show the xc hole normalized by $n_0$, $g(r)-1$, at the highest ($r_s=5$) and lowest ($r_s=90$) densities in the liquid phase, and the dashed circles mark the peak position of the real-space xc kernel $f_{\mathrm{xc}}(r)$ (see text).}
    \label{fig:sro_lro}
\end{figure}

Before turning to our analysis, we characterize how the electron liquid evolves as the coupling increases until it eventually crystallizes into a bcc Wigner crystal at $r_s\simeq87$--$113$~\cite{AZADI2022,CEPERLEY1980,DRUMMOND2004,HOLZMANN2020}. Figure~\ref{fig:sro_lro} shows the pair correlation function $g(r)$ from weak coupling up to near crystallization ($r_s=90$), obtained from separate DMC simulations of the unperturbed electron gas with a liquid trial wave function. For comparison, we also show $g(r)$ of the bcc Wigner crystal at $r_s=110$ (solid black line), taken from the DMC calculations of Ref.~\cite{DRUMMOND2004}. As the coupling increases, a peak develops in $g(r)$ and becomes pronounced for $r_s \gtrsim 20$. The location of this peak is $r\simeq 3.3/k_F$, which corresponds to the typical separation between an electron and its nearest neighbors. Interestingly, this peak coincides with the nearest-neighbor distance in the bcc Wigner crystal, $r_\mathrm{nn} = 3.38/k_F$~\footnote{The height of this peak increases with $r_s$ and then saturates, with the $r_s=80$ and $90$ curves almost on top of each other, i.e., the local order saturates before long-range order sets in.}. While the oscillations of $g(r)$ decay in the liquid, they persist to long range in the crystal phase.

The emergence of such a peak at the nearest-neighbor distance has been identified as short-range order in the literature~\cite{DAVOUDI2003,GATTOBIGIO2005}, and indicates that the liquid is approaching the Wigner crystallization transition~\cite{SMITH2024,PESCIA2024}. Apart from the peak and oscillations of $g(r)$, short-range order also shows up in the steepness of $g(r)$ on its rising edge, around $k_F r = 2$, close to the radius of the sphere available to each electron, i.e., the Wigner--Seitz radius $r_s\simeq1.92/k_F$. Compared with the weakly coupled liquid at $r_s=5$, $g(r)$ drops much more steeply toward $r \to 0$ when short-range order is present. This depletion is more easily seen in the xc hole, $n_0[g(r)-1]$, which measures the depletion of electrons around a reference electron~\cite{GIORGI2002,GIULIANI2005}. The inset of Fig.~\ref{fig:sro_lro} shows the xc hole as a two-dimensional cut through the reference electron for $r_s=5$ and $r_s=90$. The liquid with short-range order has a noticeably more localized xc hole than the weakly coupled liquid. This can be quantified by the fraction of the missing electron within the Wigner--Seitz radius, $-n_0\int_0^{r_s}[g(r)-1]\,4\pi r^2\,dr$, which would equal unity for a completely empty Wigner--Seitz sphere. This fraction grows from $57\%$ at $r_s=5$ to $95\%$ at $r_s=90$ ($98\%$ in the crystal). The localized xc hole and the nearest-neighbor peak thus appear together, suggesting that both are signatures of short-range order: the region around each electron is nearly depleted up to its Wigner--Seitz radius, while its neighbors accumulate at the nearest-neighbor distance, reminiscent of the local structure of the bcc Wigner crystal.

In reciprocal space, short-range order mainly corresponds to a peak in the static structure factor $S(q)$, arising from the oscillations in $g(r)$, followed by further oscillations, arising from the steepness of the xc hole, that are damped out at large wave vectors. Just as for $g(r)$, this peak in $S(q)$ coincides with the first Bragg peak of the incipient Wigner crystal, located at $q\simeq 2.28k_F$. The location of the peak therefore tracks the evolution from short-range order in the liquid to long-range order in the crystal. In the following, we relate these signatures of short-range order, the nearest-neighbor peak in $g(r)$ and $S(q)$ and the steepness of the xc hole, to the local field factor $G(q)$.

\subsubsection*{Calculations of $S(q)$ and $g(r)$ from $G(q)$}
The static structure factor $S(q)$ can be obtained directly from DMC. Alternatively, it follows from linear response theory via $S(q) = -\frac{1}{\pi n_0}\int_0^\infty \mathrm{Im}\,\chi(q,\omega)\,d\omega$, which, however, requires the dynamic density response function $\chi(q,\omega)$. The latter is determined by the dynamic local field factor $G(q,\omega)$, the frequency-dependent generalization of $G(q)$. Here, we use the adiabatic (static) approximation $G(q,\omega)\approx G(q)$, which is exact at $\omega=0$ and neglects the frequency dependence. Our parametrization of $G(q)$ then yields $\chi(q,\omega)$ via the Dyson equation,

\begin{equation}\label{eq:dyson_omega}
    \chi(q,\omega) = \frac{\chi_0(q,\omega)}{1-v_c(q)[1-G(q)]\chi_0(q,\omega)},
\end{equation}
where $\chi_0(q,\omega)$ is the noninteracting density response function~\cite{GIULIANI2005}. From here, one can obtain the dynamic structure factor as $S(q,\omega) = -\frac{1}{\pi n_0} \mathrm{Im}\chi(q,\omega)$.

\begin{figure}[t]
    \centering
    \includegraphics[width=\linewidth]{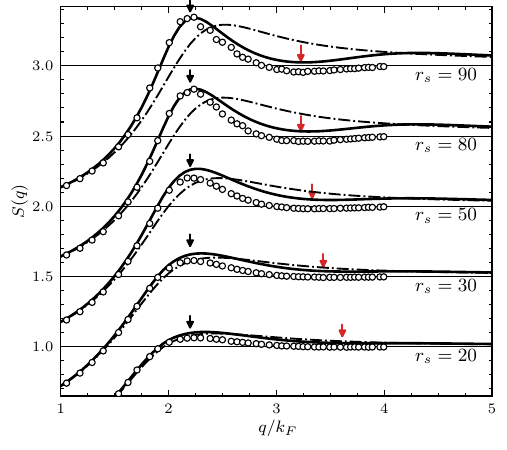}
    \caption{Static structure factor $S(q)$ at $r_s=20,30,50,80,90$ obtained from DMC calculations (circles, $N_e=162$, radially averaged) and from the adiabatic approximation (lines). The solid lines are obtained using the $G(q)$ fit from this work, while the dash-dotted lines are obtained using the $G(q)$ parametrization of Corradini \textit{et al.} Black and red arrows mark the peak and the dip of the DMC data points, respectively. Curves for successive $r_s$ are offset by $0.5$ for clarity.}
    \label{fig:sk_analysis}
\end{figure}

\begin{figure}[t]
    \centering
    \includegraphics[width=\linewidth]{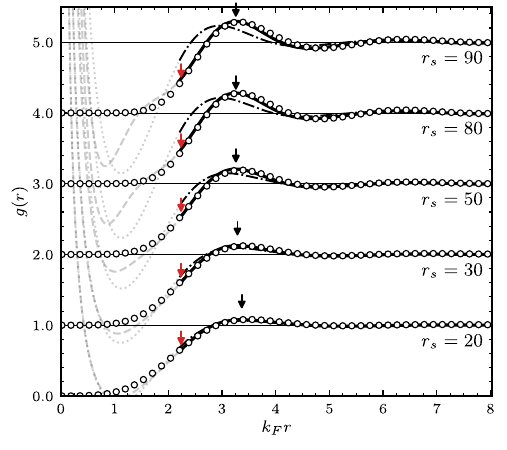}
    \caption{Pair correlation function $g(r)$ from DMC ($N_e=162$, circles) and from the adiabatic approximation (lines), obtained by Fourier transforming the corresponding $S(q)$ of Fig.~\ref{fig:sk_analysis}, with the same line styles. Gray segments mark the region where the adiabatic approximation is unreliable (see text). Red arrows mark the DMC $g(r)$ at $k_F r=2.24$, and black arrows mark the peak of $g(r)$ obtained with the $G(q)$ fit from this work. Curves for successive $r_s$ are offset by $1$ for clarity.}
    \label{fig:gofr_analysis}
\end{figure}

Figures~\ref{fig:sk_analysis} and~\ref{fig:gofr_analysis} show the static structure factor $S(q)$ and the pair correlation function $g(r)$ obtained from DMC calculations (circles, calculated independently from the unperturbed gas) and from the adiabatic approximation (lines) using Eq.~\eqref{eq:dyson_omega} for $r_s=20,30,50,80,90$. Since our goal is to investigate the effect of the feature in $G(q)$ on $S(q)$, we also show the case where $G(q)$ is featureless for comparison, i.e., the parametrization of Corradini \textit{et al.} (dash-dotted lines). In all cases, $g(r)$ is obtained from the corresponding $S(q)$ by Fourier transform. Since the adiabatic approximation neglects the frequency dependence of $G(q,\omega)$, the resulting large-$q$ behavior of $S(q)$ is incorrect, and the corresponding small-$r$ behavior of $g(r)$ is unreliable (gray curves in Fig.~\ref{fig:gofr_analysis}). In the DMC data, a peak develops in $S(q)$ near $q\simeq 2.2k_F$ as the liquid becomes more strongly coupled. As correlations grow, the peak becomes more pronounced and is followed by a dip, consistent with the short-range order discussed above.

The solid curves in Fig.~\ref{fig:sk_analysis} show that the effect of the feature in the local field factor $G(q)$ is twofold. First, it reproduces the dip in $S(q)$ after the main peak as $r_s$ increases, which a featureless $G(q)$ completely misses (red arrows). The equivalent picture in real space is shown in Fig.~\ref{fig:gofr_analysis}, where the red arrows mark the corresponding suppression of $g(r)$ on its rising edge, at $k_F r \simeq 2.24$, due to the feature in $G(q)$. Since this suppression is tied to the localization of the xc hole (Fig.~\ref{fig:sro_lro}), the feature in $G(q)$ is a signature of short-range order in the electron liquid. Consistently, the real-space xc kernel $f_{\mathrm{xc}}(r)$, i.e., the Fourier transform of $-v_c(q)G(q)$, has a sharp positive peak at $k_F r \simeq 2.2$, on the rising edge of the xc hole (dashed circles in the inset of Fig.~\ref{fig:sro_lro}), at the same distance as the red arrows in Fig.~\ref{fig:gofr_analysis}. This peak implies that density modulations cost exchange-correlation energy at this distance. Both this peak and the suppression marked by the red arrows disappear when the feature is removed from $G(q)$. For details, see the companion paper~\cite{GUNESprl}.

Second, the feature also reproduces the correct location of the main peak in $S(q)$. Without it, the main peak position is overestimated at all densities, which in real space (Fig.~\ref{fig:gofr_analysis}) corresponds to an underestimation of the typical separation between an electron and its nearest neighbors (black arrows in Figs.~\ref{fig:sk_analysis} and~\ref{fig:gofr_analysis}). These observations indicate that the features in $S(q)$, $g(r)$, and $G(q)$ are related and emerge together with short-range order. This is consistent with studies of the strongly coupled one-component plasma, the classical analog of our system, where the relation between $G(q)$ and $S(q)$ is well established~\cite{Ichimaru2004}. Although the adiabatic approximation is quantitatively inaccurate at large $q$ and small $r$, as discussed above, the qualitative agreement with DMC is clear. The same agreement also holds for other approximations, such as the single-mode approximation (SMA), in which the dynamic structure factor is approximated by a single sharp collective mode, $S(q,\omega)\approx S(q)\delta(\omega-\omega_q)$~\cite{GUNESprl}.

\section{Conclusions and Outlook}\label{sec:conclusion}
We have presented diffusion Monte Carlo calculations of the static density response function $\chi(q)$ and the local field factor $G(q)$ of the three-dimensional electron gas at zero temperature. Our results extend the zero-temperature QMC data beyond the metallic regime and across the liquid phase up to the vicinity of Wigner crystallization. These calculations were made possible by a careful analysis and control of the systematic errors that arise in the strongly coupled regime. First, we established a semi-empirical scaling of the linear-response window with density, which guides the choice of $v_\mathbf q$. Second, we demonstrated that a backflow transformation, while lowering the energy substantially, leaves the extracted $\chi(q)$ unchanged, so that Slater--Jastrow trial wave functions suffice for the response. Third, we showed that the scaling $\alpha$ of the perturbing potential entering the Slater determinant must be optimized, since a poor choice alters the curvature of $E(v_\mathbf{q})$ and can even reverse its sign and thereby lead to an unphysical response. We provided a closed-form expression for the optimal $\alpha(q,r_s)$, valid for $2\le r_s\le100$, which removes the need for a separate optimization in future calculations.

Finally, we showed that the local field factor remains insensitive to system size even at very large $r_s$ values wherever our data resolve it, and that the correction based on the noninteracting response alone removes the dominant finite-size error in $\chi(q)$. The time step was chosen small enough to ensure that the time-step bias is negligible. Together, these results show that the strongly coupled regime can be treated with standard Slater--Jastrow wave functions and modest system sizes, paving the way for future calculations of the response of the electron gas and related systems.

At metallic densities, our results agree with previous DMC data and parametrizations, extend them to $q<k_F$, where they are consistent with the compressibility sum rule, and resolve the discrepancy between DMC and diagrammatic QMC at intermediate wave vectors. Beyond this regime, $G(q)$ develops a pronounced nonmonotonic structure at intermediate wave vectors, already discernible at $r_s=5$, whose amplitude grows with $r_s$ and which is absent from all existing parametrizations. The quantitative agreement with finite-temperature PIMC data at $r_s=20$--$50$ establishes this feature as a ground-state property of the electron liquid. Finally, we have shown that the inclusion of this feature in $G(q)$ reproduces the correct physical behaviors of the pair correlation function $g(r)$ and the static structure factor $S(q)$, which are missed by a featureless $G(q)$, and that this feature is related to the emergence of short-range order in the electron liquid, as seen in the calculations of $g(r)$ and $S(q)$.

The parametrization of $G(q)$ used throughout this work, given in the companion paper~\cite{GUNESprl}, reproduces our data over the full liquid phase and is directly usable for numerous applications. For instance, it can be used to construct exchange-correlation functionals in time-dependent density functional theory~\cite{Runge1984,Petersilka1996,Ullrich2012}, which is widely used for electronic structure calculations of materials. It can also be used to describe dynamical and spectral properties in the electron gas~\cite{PANHOLZER2018,KOSKELO2025} and in real materials~\cite{PANHOLZER2018,CAZZANIGA2011,Chulkov2006,Pitarke2003,Patrick2015,Olsen2019}, to construct effective electron--electron interactions~\cite{KUKKONEN1979,TAKADA1993,DORNHEIM2022b}, and to extend response calculations of warm dense matter to lower temperatures~\cite{DORNHEIM2019}, where current PIMC calculations are limited by the fermion sign problem. Additionally, our results can serve as benchmarks for future theoretical and computational studies of the electron gas and related systems.

The methodology developed here is not specific to the density response of the unpolarized three-dimensional gas. In fact, it can be applied directly to the spin response, through a spin-dependent perturbation, to partially polarized systems, and to the two-dimensional electron gas, where crystallization occurs at much higher density and where the interplay between charge and spin correlations at intermediate wave vectors remains to be established.

\section*{Data Availability}
The data that support the findings of this study, i.e., the local field factor $G(q)$ from which the static density response function $\chi(q)$ can be obtained, the static structure factor $S(q)$, and the pair correlation function $g(r)$, are openly available in the Zenodo repository~\cite{zenodo}.

\begin{acknowledgments}
    The authors would like to thank Saverio Moroni, Carl Kukkonen, and Carlos Rodriguez Perez for helpful discussions. The Flatiron Institute is a division of the Simons Foundation. This work has received state funding managed by the French National Research Agency (ANR) under the France 2030 program, reference ``Grant No.~ANR-22-EXES-0013''. YY acknowledges support by the National Science Foundation (NSF) under grant number DMR-2532734. This work used Delta through allocation PHY250273 from the Advanced Cyberinfrastructure Coordination Ecosystem: Services \& Support (ACCESS) program, which is supported by U.S. National Science Foundation grants \#2138259, \#2138286, \#2138307, \#2137603, and \#2138296.
    This research used the Delta advanced computing and data resource which is supported by the National Science Foundation (award OAC 2005572) and the State of Illinois. Delta is a joint effort of the University of Illinois Urbana-Champaign and its National Center for Supercomputing Applications.
\end{acknowledgments}

\appendix
\section{Low- and high-\texorpdfstring{$q$}{q} limits of \texorpdfstring{$G(q)$}{G(q)}}\label{app:limits}

We use Hartree atomic units, $k_F = (9\pi/4)^{1/3}/r_s$ and $n_0 = 3/(4\pi r_s^3)$. The local field factor satisfies
\begin{equation}
    G(q) \xrightarrow{q\to0} A\left(\frac{q}{k_F}\right)^2, \qquad
    G(q) \xrightarrow{q\to\infty} C\left(\frac{q}{k_F}\right)^2 + B .
\end{equation}

The coefficient $A$ follows from the compressibility sum rule~\cite{MORONI1995,CORRADINI1998},
\begin{equation}\label{eq:A_limit}
    \begin{split}
        A(r_s) & = \frac{1}{4} - \frac{k_F^2}{4\pi}\frac{d\mu_c}{dn}                                              \\
               & = \frac{1}{4} - \frac{(9\pi/4)^{2/3}\,r_s^2}{27}\left(r_s\,\epsilon_c'' - 2\,\epsilon_c'\right),
    \end{split}
\end{equation}
where $\mu_c = d(n\epsilon_c)/dn$, primes denote derivatives with respect to $r_s$, and the exchange contribution is the constant $1/4$.

The coefficient $C$ is set by the correlation contribution to the kinetic energy~\cite{HOLAS1987,MORONI1995},
\begin{equation}\label{eq:C_limit}
    C(r_s) = -\frac{\pi}{2k_F}\frac{d(r_s\epsilon_c)}{dr_s} = \frac{\pi}{2k_F}\,t_c ,
\end{equation}
where $t_c = -\epsilon_c - r_s\epsilon_c'$ is the correlation kinetic energy per electron given by the virial theorem.

For the correlation energy per electron $\epsilon_c(r_s)$ we use the Perdew--Wang parametrization~\cite{PERDEW1992} of the unpolarized gas,
\begin{equation}\label{eq:pw92}
    \epsilon_c(r_s) = -2a_0^\mathrm{PW}(1+\alpha_1^\mathrm{PW} r_s)\ln\!\left(1 + \frac{1}{D}\right),
\end{equation}
where $D = 2a_0^\mathrm{PW}\left(\beta_1 r_s^{1/2} + \beta_2 r_s + \beta_3 r_s^{3/2} + \beta_4 r_s^2\right)$, with $a_0^\mathrm{PW} = 0.031091$, $\alpha_1^\mathrm{PW} = 0.21370$, $\beta_1 = 7.5957$, $\beta_2 = 3.5876$, $\beta_3 = 1.6382$, and $\beta_4 = 0.49294$. The same $\epsilon_c$ enters both $A$ and $C$, and we use it for every model of $G(q)$ shown in this work, including those of Refs.~\cite{MORONI1995,CORRADINI1998}. Evaluating $A$ and $C$ with the Perdew--Zunger form~\cite{PERDEW1981} instead changes $A$ by less than $0.5\%$ and $C$ by $1$--$4\%$. The resulting change in $\chi(q)$ is below $0.3\%$ for $r_s\leq20$, and reaches about one statistical error bar at $r_s=90$.

The constant $B$ is taken from the parametrization of Ref.~\cite{MORONI1995}, which is based on the DMC calculations of Ref.~\cite{ORTIZ1994},
\begin{equation}\label{eq:B_limit}
    B(r_s) = \frac{1 + a_1^B r_s^{1/2} + a_2^B r_s^{3/2}}{3 + b_1^B r_s^{1/2} + b_2^B r_s^{3/2}},
\end{equation}
with $a_1^B = 2.15$, $a_2^B = 0.435$, $b_1^B = 1.57$, and $b_2^B = 0.409$. Equation~\eqref{eq:B_limit} was fitted for $r_s\leq10$, so for $r_s>10$ it is an extrapolation, and is not analyzed in this work since the regime it affects is either not covered by our DMC data or the error bars are too large to be sensitive to it.

\section{Determining the perturbation range}\label{app:vq_range}
In this appendix, we describe our procedure for selecting the range of perturbation amplitudes $v_\mathbf{q}$ used to fit the DMC energies $E(v_\mathbf{q})$ and extract $\chi(q)$, which is crucial for obtaining reliable results. The main challenge is to select a range that is sufficiently small to remain in the linear-response regime, but sufficiently large to ensure a good signal-to-noise ratio (SNR) in the DMC energies. At small $v_\mathbf{q}$, the energy shift $E(v_\mathbf{q})-E_0\propto v_\mathbf{q}^2$ becomes comparable to the statistical error of the DMC energy, so the curvature is poorly determined, which leads to very large error bars in $\chi(q)$. At large $v_\mathbf{q}$, the higher-order terms in Eq.~\eqref{eqn:expansion} become significant and bias the quadratic coefficient, and hence $\chi(q)$. Since each amplitude requires an independent optimization and DMC calculation, a poorly chosen range is also costly. We have found that this range depends on both the density and the wave vector, and we have developed a systematic procedure to determine it. Over the densities considered here, the appropriate amplitudes span almost two orders of magnitude.

\begin{figure}[b]
    \centering
    \includegraphics[width=\linewidth]{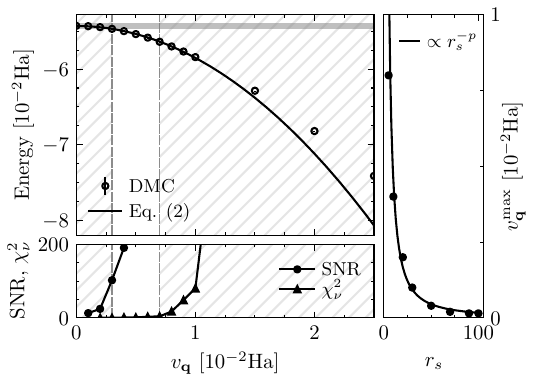}
    \caption{Selection of the $v_\mathbf{q}$ fitting range for $r_s=10$, $N_e=54$, $q/k_F=1.61$. Main panel: DMC energies (open circles) with the quadratic fit of Eq.~\eqref{eqn:expansion} (solid). The gray band is the error bar of $E_0$ magnified $100\times$. Lower panel: SNR (circles) and reduced $\chi^2_\nu$ of the quadratic fit (triangles). Hatched areas mark the excluded low-SNR (small $v_\mathbf{q}$) and nonlinear (large $v_\mathbf{q}$) regions, and dashed lines delimit the linear-response window. Right panel: maximum $v_\mathbf{q}$ used in this work vs.\ $r_s$, with fit $v_\mathbf{q}^\mathrm{max} = 0.1212\,r_s^{-1.43}$~Ha (solid line).}
    \label{fig:vq_range}
\end{figure}

Figure~\ref{fig:vq_range} demonstrates our strategy for selecting the range of $v_\mathbf{q}$ used to fit $E(v_\mathbf{q})$ and extract $\chi(q)$, for a single density and wave vector. The main panel shows the DMC energies (symbols) with the quadratic fit of Eq.~\eqref{eqn:expansion} (solid). As one can see, one needs to select a range of $v_\mathbf{q}$ that is neither too small, where the SNR is low, nor too large, where the response is no longer linear. We indicate these two regions with hatched areas in the main panel.  The lower panel shows the SNR (circles) and reduced $\chi^2_\nu$ of the quadratic fit (triangles). The best extraction of $\chi(q)$ is obtained in the linear-response regime, where the SNR is sufficiently high and the reduced $\chi^2_\nu$ of the quadratic fit is sufficiently low. Beyond the linear regime, $\chi^2_\nu$ rises by one to three orders of magnitude within a few $v_\mathbf{q}$ points (see lower panel), so any threshold of order unity selects the same range. We have used the criteria of SNR $> 10$ and $\chi^2_\nu < 5$ to select the range of $v_\mathbf{q}$ used for the quadratic fit.

The demonstration in Fig.~\ref{fig:vq_range} is done for a single point. Selecting the valid range of $v_\mathbf{q}$, on the other hand, has a nontrivial dependence on both $r_s$ and $q$. As we have seen in Figs.~\ref{fig:backflow}(a) and~\ref{fig:backflow}(b), the linear-response regime shrinks significantly as one goes from $r_s=5$ to $r_s=70$. Determining this dependence by brute force is prohibitively expensive. For this, we have solved the noninteracting problem perturbed by the same cosine potential, and found that the maximum $v_\mathbf{q}$ allowed to remain in the quadratic regime scales as $r_s^{-p}$ with $p\simeq1.4$--$1.5$. Since we have found the interacting linear-response window to be smaller than the noninteracting one, we then determined the interacting $v_\mathbf{q}^\mathrm{max}$ at a single density ($r_s=5$), by sampling $v_\mathbf{q}$ between zero and the noninteracting maximum, and we transferred it to the other densities assuming the same $r_s^{-p}$ scaling. In each case, $E(v_\mathbf{q})$ was verified to remain in the quadratic regime (by the criteria defined above). The resulting amplitudes are shown in the right panel of Fig.~\ref{fig:vq_range}, together with the fit $v_\mathbf{q}^\mathrm{max} = 0.1212\,r_s^{-1.43}$~Ha, whose exponent is taken from the noninteracting problem and whose prefactor is fitted to the data points. This avoids scanning $v_\mathbf{q}$ at every density, and it provides a guide for choosing the perturbation amplitude in future calculations. Note that the window also depends on $q$: around $q\sim 2k_F$, the linear-response regime for higher $r_s$ values is significantly smaller compared to other wave vectors, as shown by the dash-dotted lines in Figs.~\ref{fig:backflow}(a) and~\ref{fig:backflow}(b). We handle this by successively discarding the largest perturbation amplitude until the criteria defined above, i.e., SNR $> 10$ and $\chi^2_\nu < 5$, hold. All results that we present in this work are obtained using this procedure and criteria.

\section{Effects of twist averaging on $\chi(q)$}\label{app:twist}

\begin{figure}[t]
    \centering
    \includegraphics[width=\linewidth]{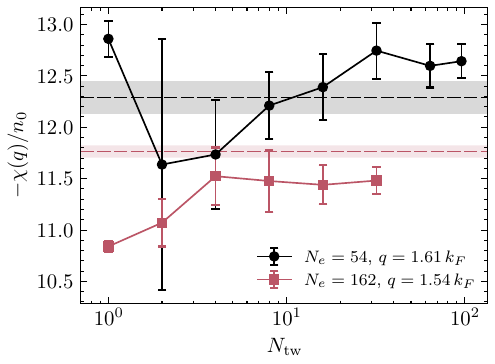}
    \caption{Twist-averaged static response at $r_s=5$ as a function of the number of twists $N_\mathrm{tw}$, for $N_e=54$ ($q=1.61\,k_F$) and $N_e=162$ ($q=1.54\,k_F$). Symbols: fixed-$N_e$ twist averages without finite-size correction. Dashed lines: $\Gamma$-point result corrected with Eq.~\eqref{eq:fse}, with shaded bands indicating its statistical uncertainty. }
    \label{fig:twist_fse}
\end{figure}

Twist-averaged boundary conditions~\cite{Lin_twist-averaged-boundary_2001} are a common alternative to analytic corrections for one-body finite-size errors. We test them at $r_s=5$ for $N_e=54$ with $\mathbf q=(3,0,0)\,2\pi/L$ and $N_e=162$ with $\mathbf q=(3,2,2)\,2\pi/L$. For each twist, drawn from a Halton sequence starting at $\Gamma$, we perform DMC calculations at fixed $N_e$ and at the production perturbation amplitudes, average $E(v_\mathbf{q})$ over twists, and extract $\chi(q)$ from its curvature. Figure~\ref{fig:twist_fse} shows the result as a function of the number of twists, together with the $\Gamma$-point value corrected with Eq.~\eqref{eq:fse}. With increasing $N_\mathrm{tw}$, the twist average moves away from the uncorrected $\Gamma$ value ($N_\mathrm{tw}=1$) toward the corrected one, recovering most of the correction. For $N_e=162$, $-\chi/n_0$ goes from $10.84$ at $\Gamma$ to $11.48\pm0.13$ after 32 twists, compared with $11.76\pm0.05$ for the corrected $\Gamma$ value. For $N_e=54$, the uncorrected $\Gamma$ value is already closer to the corrected one, and the twist average remains within $2\sigma$ of it for all $N_\mathrm{tw}\geq2$, reaching $12.64\pm0.16$ after 96 twists, compared with $12.29\pm0.16$. In both cases, the twist average agrees with the corrected $\Gamma$ value to within 2--3\%, i.e., within about $2\sigma$. The remaining difference is expected for fixed-$N_e$ twist averaging. For a size-independent $G(q)$ [Fig.~\ref{fig:fse}(a)], the twist dependence of $\chi(q)$ enters only through the noninteracting response at each twist, so the procedure can reach the thermodynamic limit only if it recovers the Lindhard function for noninteracting electrons. At fixed $N_e$ it does not, because each twist fills its own $N_e$ lowest levels and the averaged Fermi sphere is distorted~\cite{Lin_twist-averaged-boundary_2001}. Fixing the chemical potential instead avoids this at the cost of a fluctuating particle number (grand-canonical twist averaging), which could in principle lead to a more accurate size correction but was not tested here. Throughout this work, we keep $N_e$ fixed for all twists (canonical twist averaging) and use Eq.~\eqref{eq:fse}, which inserts the thermodynamic-limit $\chi_0$ directly from a single twist.

\bibliographystyle{apsrev4-2}
\bibliography{main}

\begin{thebibliography}{88}%
\makeatletter
\providecommand \@ifxundefined [1]{%
 \@ifx{#1\undefined}
}%
\providecommand \@ifnum [1]{%
 \ifnum #1\expandafter \@firstoftwo
 \else \expandafter \@secondoftwo
 \fi
}%
\providecommand \@ifx [1]{%
 \ifx #1\expandafter \@firstoftwo
 \else \expandafter \@secondoftwo
 \fi
}%
\providecommand \natexlab [1]{#1}%
\providecommand \enquote  [1]{``#1''}%
\providecommand \bibnamefont  [1]{#1}%
\providecommand \bibfnamefont [1]{#1}%
\providecommand \citenamefont [1]{#1}%
\providecommand \href@noop [0]{\@secondoftwo}%
\providecommand \href [0]{\begingroup \@sanitize@url \@href}%
\providecommand \@href[1]{\@@startlink{#1}\@@href}%
\providecommand \@@href[1]{\endgroup#1\@@endlink}%
\providecommand \@sanitize@url [0]{\catcode `\\12\catcode `\$12\catcode `\&12\catcode `\#12\catcode `\^12\catcode `\_12\catcode `\%12\relax}%
\providecommand \@@startlink[1]{}%
\providecommand \@@endlink[0]{}%
\providecommand \url  [0]{\begingroup\@sanitize@url \@url }%
\providecommand \@url [1]{\endgroup\@href {#1}{\urlprefix }}%
\providecommand \urlprefix  [0]{URL }%
\providecommand \Eprint [0]{\href }%
\providecommand \doibase [0]{https://doi.org/}%
\providecommand \selectlanguage [0]{\@gobble}%
\providecommand \bibinfo  [0]{\@secondoftwo}%
\providecommand \bibfield  [0]{\@secondoftwo}%
\providecommand \translation [1]{[#1]}%
\providecommand \BibitemOpen [0]{}%
\providecommand \bibitemStop [0]{}%
\providecommand \bibitemNoStop [0]{.\EOS\space}%
\providecommand \EOS [0]{\spacefactor3000\relax}%
\providecommand \BibitemShut  [1]{\csname bibitem#1\endcsname}%
\let\auto@bib@innerbib\@empty
\bibitem [{\citenamefont {Kugler}(1975)}]{KUGLER1975}%
  \BibitemOpen
  \bibfield  {author} {\bibinfo {author} {\bibfnamefont {A.~A.}\ \bibnamefont {Kugler}},\ }\href {https://doi.org/10.1007/BF01024183} {\bibfield  {journal} {\bibinfo  {journal} {Journal of Statistical Physics}\ }\textbf {\bibinfo {volume} {12}},\ \bibinfo {pages} {35} (\bibinfo {year} {1975})}\BibitemShut {NoStop}%
\bibitem [{\citenamefont {Singwi}\ and\ \citenamefont {Tosi}(1981)}]{SINGWITOSI1981}%
  \BibitemOpen
  \bibfield  {author} {\bibinfo {author} {\bibfnamefont {K.~S.}\ \bibnamefont {Singwi}}\ and\ \bibinfo {author} {\bibfnamefont {M.~P.}\ \bibnamefont {Tosi}},\ }in\ \href {https://doi.org/10.1016/S0081-1947(08)60116-2} {\emph {\bibinfo {booktitle} {Solid State Physics}}},\ Vol.~\bibinfo {volume} {36},\ \bibinfo {editor} {edited by\ \bibinfo {editor} {\bibfnamefont {H.}~\bibnamefont {Ehrenreich}}, \bibinfo {editor} {\bibfnamefont {F.}~\bibnamefont {Seitz}},\ and\ \bibinfo {editor} {\bibfnamefont {D.}~\bibnamefont {Turnbull}}}\ (\bibinfo  {publisher} {Academic Press},\ \bibinfo {year} {1981})\ pp.\ \bibinfo {pages} {177--266}\BibitemShut {NoStop}%
\bibitem [{\citenamefont {Giuliani}\ and\ \citenamefont {Vignale}(2005)}]{GIULIANI2005}%
  \BibitemOpen
  \bibfield  {author} {\bibinfo {author} {\bibfnamefont {G.}~\bibnamefont {Giuliani}}\ and\ \bibinfo {author} {\bibfnamefont {G.}~\bibnamefont {Vignale}},\ }\href {https://doi.org/10.1017/CBO9780511619915} {\emph {\bibinfo {title} {Quantum {{Theory}} of the {{Electron Liquid}}}}},\ \bibinfo {edition} {1st}\ ed.\ (\bibinfo  {publisher} {Cambridge University Press},\ \bibinfo {year} {2005})\BibitemShut {NoStop}%
\bibitem [{\citenamefont {Pines}\ and\ \citenamefont {Nozi{\`e}res}(1966)}]{PINES1966}%
  \BibitemOpen
  \bibfield  {author} {\bibinfo {author} {\bibfnamefont {D.}~\bibnamefont {Pines}}\ and\ \bibinfo {author} {\bibfnamefont {P.}~\bibnamefont {Nozi{\`e}res}},\ }\href@noop {} {\emph {\bibinfo {title} {The Theory of Quantum Liquids, {{Vol}}. {{I}}: {{Normal Fermi Liquids}}}}}\ (\bibinfo  {publisher} {W. A. Benjamin},\ \bibinfo {address} {New York},\ \bibinfo {year} {1966})\BibitemShut {NoStop}%
\bibitem [{\citenamefont {Kohn}\ and\ \citenamefont {Sham}(1965)}]{KOHN1965a}%
  \BibitemOpen
  \bibfield  {author} {\bibinfo {author} {\bibfnamefont {W.}~\bibnamefont {Kohn}}\ and\ \bibinfo {author} {\bibfnamefont {L.~J.}\ \bibnamefont {Sham}},\ }\href {https://doi.org/10.1103/PhysRev.137.A1697} {\bibfield  {journal} {\bibinfo  {journal} {Physical Review}\ }\textbf {\bibinfo {volume} {137}},\ \bibinfo {pages} {A1697} (\bibinfo {year} {1965})}\BibitemShut {NoStop}%
\bibitem [{\citenamefont {Ichimaru}(1982)}]{ICHIMARU1982}%
  \BibitemOpen
  \bibfield  {author} {\bibinfo {author} {\bibfnamefont {S.}~\bibnamefont {Ichimaru}},\ }\href {https://doi.org/10.1103/RevModPhys.54.1017} {\bibfield  {journal} {\bibinfo  {journal} {Reviews of Modern Physics}\ }\textbf {\bibinfo {volume} {54}},\ \bibinfo {pages} {1017} (\bibinfo {year} {1982})}\BibitemShut {NoStop}%
\bibitem [{\citenamefont {Kukkonen}\ and\ \citenamefont {Overhauser}(1979)}]{KUKKONEN1979}%
  \BibitemOpen
  \bibfield  {author} {\bibinfo {author} {\bibfnamefont {C.~A.}\ \bibnamefont {Kukkonen}}\ and\ \bibinfo {author} {\bibfnamefont {A.~W.}\ \bibnamefont {Overhauser}},\ }\href {https://doi.org/10.1103/PhysRevB.20.550} {\bibfield  {journal} {\bibinfo  {journal} {Physical Review B}\ }\textbf {\bibinfo {volume} {20}},\ \bibinfo {pages} {550} (\bibinfo {year} {1979})}\BibitemShut {NoStop}%
\bibitem [{\citenamefont {Kukkonen}\ and\ \citenamefont {Chen}(2021)}]{KUKKONEN2021}%
  \BibitemOpen
  \bibfield  {author} {\bibinfo {author} {\bibfnamefont {C.~A.}\ \bibnamefont {Kukkonen}}\ and\ \bibinfo {author} {\bibfnamefont {K.}~\bibnamefont {Chen}},\ }\href {https://doi.org/10.1103/PhysRevB.104.195142} {\bibfield  {journal} {\bibinfo  {journal} {Physical Review B}\ }\textbf {\bibinfo {volume} {104}},\ \bibinfo {pages} {195142} (\bibinfo {year} {2021})}\BibitemShut {NoStop}%
\bibitem [{\citenamefont {Panholzer}\ \emph {et~al.}(2018)\citenamefont {Panholzer}, \citenamefont {Gatti},\ and\ \citenamefont {Reining}}]{PANHOLZER2018}%
  \BibitemOpen
  \bibfield  {author} {\bibinfo {author} {\bibfnamefont {M.}~\bibnamefont {Panholzer}}, \bibinfo {author} {\bibfnamefont {M.}~\bibnamefont {Gatti}},\ and\ \bibinfo {author} {\bibfnamefont {L.}~\bibnamefont {Reining}},\ }\href {https://doi.org/10.1103/PhysRevLett.120.166402} {\bibfield  {journal} {\bibinfo  {journal} {Physical Review Letters}\ }\textbf {\bibinfo {volume} {120}},\ \bibinfo {pages} {166402} (\bibinfo {year} {2018})}\BibitemShut {NoStop}%
\bibitem [{\citenamefont {Hubbard}(1958)}]{HUBBARD1958}%
  \BibitemOpen
  \bibfield  {author} {\bibinfo {author} {\bibfnamefont {J.}~\bibnamefont {Hubbard}},\ }\href {https://doi.org/10.1098/rspa.1958.0003} {\bibfield  {journal} {\bibinfo  {journal} {Proceedings of the Royal Society of London. A. Mathematical and Physical Sciences}\ }\textbf {\bibinfo {volume} {243}},\ \bibinfo {pages} {336} (\bibinfo {year} {1958})}\BibitemShut {NoStop}%
\bibitem [{\citenamefont {Geldart}\ and\ \citenamefont {Vosko}(1966)}]{GELDART1966}%
  \BibitemOpen
  \bibfield  {author} {\bibinfo {author} {\bibfnamefont {D.~J.~W.}\ \bibnamefont {Geldart}}\ and\ \bibinfo {author} {\bibfnamefont {S.~H.}\ \bibnamefont {Vosko}},\ }\href {https://doi.org/10.1139/p66-174} {\bibfield  {journal} {\bibinfo  {journal} {Canadian Journal of Physics}\ }\textbf {\bibinfo {volume} {44}},\ \bibinfo {pages} {2137} (\bibinfo {year} {1966})},\ \Eprint {https://arxiv.org/abs/https://doi.org/10.1139/p66-174} {https://doi.org/10.1139/p66-174} \BibitemShut {NoStop}%
\bibitem [{\citenamefont {Singwi}\ \emph {et~al.}(1968)\citenamefont {Singwi}, \citenamefont {Tosi}, \citenamefont {Land},\ and\ \citenamefont {Sj{\"o}lander}}]{SINGWI1968}%
  \BibitemOpen
  \bibfield  {author} {\bibinfo {author} {\bibfnamefont {K.~S.}\ \bibnamefont {Singwi}}, \bibinfo {author} {\bibfnamefont {M.~P.}\ \bibnamefont {Tosi}}, \bibinfo {author} {\bibfnamefont {R.~H.}\ \bibnamefont {Land}},\ and\ \bibinfo {author} {\bibfnamefont {A.}~\bibnamefont {Sj{\"o}lander}},\ }\href {https://doi.org/10.1103/PhysRev.176.589} {\bibfield  {journal} {\bibinfo  {journal} {Physical Review}\ }\textbf {\bibinfo {volume} {176}},\ \bibinfo {pages} {589} (\bibinfo {year} {1968})}\BibitemShut {NoStop}%
\bibitem [{\citenamefont {Geldart}\ and\ \citenamefont {Taylor}(1970)}]{GELDART1970}%
  \BibitemOpen
  \bibfield  {author} {\bibinfo {author} {\bibfnamefont {D.~J.~W.}\ \bibnamefont {Geldart}}\ and\ \bibinfo {author} {\bibfnamefont {R.}~\bibnamefont {Taylor}},\ }\href {https://doi.org/10.1139/p70-022} {\bibfield  {journal} {\bibinfo  {journal} {Canadian Journal of Physics}\ }\textbf {\bibinfo {volume} {48}},\ \bibinfo {pages} {155} (\bibinfo {year} {1970})}\BibitemShut {NoStop}%
\bibitem [{\citenamefont {Toigo}\ and\ \citenamefont {Woodruff}(1970)}]{TOIGO1970}%
  \BibitemOpen
  \bibfield  {author} {\bibinfo {author} {\bibfnamefont {F.}~\bibnamefont {Toigo}}\ and\ \bibinfo {author} {\bibfnamefont {T.~O.}\ \bibnamefont {Woodruff}},\ }\href {https://doi.org/10.1103/PhysRevB.2.3958} {\bibfield  {journal} {\bibinfo  {journal} {Phys. Rev. B}\ }\textbf {\bibinfo {volume} {2}},\ \bibinfo {pages} {3958} (\bibinfo {year} {1970})}\BibitemShut {NoStop}%
\bibitem [{\citenamefont {Toigo}\ and\ \citenamefont {Woodruff}(1971)}]{TOIGO1971}%
  \BibitemOpen
  \bibfield  {author} {\bibinfo {author} {\bibfnamefont {F.}~\bibnamefont {Toigo}}\ and\ \bibinfo {author} {\bibfnamefont {T.~O.}\ \bibnamefont {Woodruff}},\ }\href {https://doi.org/10.1103/PhysRevB.4.371} {\bibfield  {journal} {\bibinfo  {journal} {Phys. Rev. B}\ }\textbf {\bibinfo {volume} {4}},\ \bibinfo {pages} {371} (\bibinfo {year} {1971})}\BibitemShut {NoStop}%
\bibitem [{\citenamefont {Vashishta}\ and\ \citenamefont {Singwi}(1972)}]{VASHISHTA1972}%
  \BibitemOpen
  \bibfield  {author} {\bibinfo {author} {\bibfnamefont {P.}~\bibnamefont {Vashishta}}\ and\ \bibinfo {author} {\bibfnamefont {K.~S.}\ \bibnamefont {Singwi}},\ }\href {https://doi.org/10.1103/PhysRevB.6.875} {\bibfield  {journal} {\bibinfo  {journal} {Physical Review B}\ }\textbf {\bibinfo {volume} {6}},\ \bibinfo {pages} {875} (\bibinfo {year} {1972})}\BibitemShut {NoStop}%
\bibitem [{\citenamefont {Rajagopal}(1972)}]{RAJAGOPAL1972}%
  \BibitemOpen
  \bibfield  {author} {\bibinfo {author} {\bibfnamefont {A.~K.}\ \bibnamefont {Rajagopal}},\ }\href {https://doi.org/10.1103/PhysRevA.6.1239} {\bibfield  {journal} {\bibinfo  {journal} {Phys. Rev. A}\ }\textbf {\bibinfo {volume} {6}},\ \bibinfo {pages} {1239} (\bibinfo {year} {1972})}\BibitemShut {NoStop}%
\bibitem [{\citenamefont {Tripathy}\ and\ \citenamefont {Mandal}(1977)}]{TRIPATHY1977}%
  \BibitemOpen
  \bibfield  {author} {\bibinfo {author} {\bibfnamefont {D.~N.}\ \bibnamefont {Tripathy}}\ and\ \bibinfo {author} {\bibfnamefont {S.~S.}\ \bibnamefont {Mandal}},\ }\href {https://doi.org/10.1103/PhysRevB.16.231} {\bibfield  {journal} {\bibinfo  {journal} {Phys. Rev. B}\ }\textbf {\bibinfo {volume} {16}},\ \bibinfo {pages} {231} (\bibinfo {year} {1977})}\BibitemShut {NoStop}%
\bibitem [{\citenamefont {Holas}\ \emph {et~al.}(1979)\citenamefont {Holas}, \citenamefont {Aravind},\ and\ \citenamefont {Singwi}}]{HOLAS1979}%
  \BibitemOpen
  \bibfield  {author} {\bibinfo {author} {\bibfnamefont {A.}~\bibnamefont {Holas}}, \bibinfo {author} {\bibfnamefont {P.~K.}\ \bibnamefont {Aravind}},\ and\ \bibinfo {author} {\bibfnamefont {K.~S.}\ \bibnamefont {Singwi}},\ }\href {https://doi.org/10.1103/PhysRevB.20.4912} {\bibfield  {journal} {\bibinfo  {journal} {Phys. Rev. B}\ }\textbf {\bibinfo {volume} {20}},\ \bibinfo {pages} {4912} (\bibinfo {year} {1979})}\BibitemShut {NoStop}%
\bibitem [{\citenamefont {Devreese}\ \emph {et~al.}(1980)\citenamefont {Devreese}, \citenamefont {Brosens},\ and\ \citenamefont {Lemmens}}]{DEVREESE1980}%
  \BibitemOpen
  \bibfield  {author} {\bibinfo {author} {\bibfnamefont {J.~T.}\ \bibnamefont {Devreese}}, \bibinfo {author} {\bibfnamefont {F.}~\bibnamefont {Brosens}},\ and\ \bibinfo {author} {\bibfnamefont {L.~F.}\ \bibnamefont {Lemmens}},\ }\href {https://doi.org/10.1103/PhysRevB.21.1349} {\bibfield  {journal} {\bibinfo  {journal} {Phys. Rev. B}\ }\textbf {\bibinfo {volume} {21}},\ \bibinfo {pages} {1349} (\bibinfo {year} {1980})}\BibitemShut {NoStop}%
\bibitem [{\citenamefont {Brosens}\ \emph {et~al.}(1980)\citenamefont {Brosens}, \citenamefont {Devreese},\ and\ \citenamefont {Lemmens}}]{BROSENS1980}%
  \BibitemOpen
  \bibfield  {author} {\bibinfo {author} {\bibfnamefont {F.}~\bibnamefont {Brosens}}, \bibinfo {author} {\bibfnamefont {J.~T.}\ \bibnamefont {Devreese}},\ and\ \bibinfo {author} {\bibfnamefont {L.~F.}\ \bibnamefont {Lemmens}},\ }\href {https://doi.org/10.1103/PhysRevB.21.1363} {\bibfield  {journal} {\bibinfo  {journal} {Physical Review B}\ }\textbf {\bibinfo {volume} {21}},\ \bibinfo {pages} {1363} (\bibinfo {year} {1980})}\BibitemShut {NoStop}%
\bibitem [{\citenamefont {Utsumi}\ and\ \citenamefont {Ichimaru}(1980{\natexlab{a}})}]{UTSUMI1980}%
  \BibitemOpen
  \bibfield  {author} {\bibinfo {author} {\bibfnamefont {K.}~\bibnamefont {Utsumi}}\ and\ \bibinfo {author} {\bibfnamefont {S.}~\bibnamefont {Ichimaru}},\ }\href {https://doi.org/10.1103/PhysRevB.22.5203} {\bibfield  {journal} {\bibinfo  {journal} {Physical Review B}\ }\textbf {\bibinfo {volume} {22}},\ \bibinfo {pages} {5203} (\bibinfo {year} {1980}{\natexlab{a}})}\BibitemShut {NoStop}%
\bibitem [{\citenamefont {Utsumi}\ and\ \citenamefont {Ichimaru}(1980{\natexlab{b}})}]{UTSUMI1980B}%
  \BibitemOpen
  \bibfield  {author} {\bibinfo {author} {\bibfnamefont {K.}~\bibnamefont {Utsumi}}\ and\ \bibinfo {author} {\bibfnamefont {S.}~\bibnamefont {Ichimaru}},\ }\href {https://doi.org/10.1103/PhysRevB.22.1522} {\bibfield  {journal} {\bibinfo  {journal} {Phys. Rev. B}\ }\textbf {\bibinfo {volume} {22}},\ \bibinfo {pages} {1522} (\bibinfo {year} {1980}{\natexlab{b}})}\BibitemShut {NoStop}%
\bibitem [{\citenamefont {Ichimaru}\ and\ \citenamefont {Utsumi}(1981)}]{UTSUMI1981}%
  \BibitemOpen
  \bibfield  {author} {\bibinfo {author} {\bibfnamefont {S.}~\bibnamefont {Ichimaru}}\ and\ \bibinfo {author} {\bibfnamefont {K.}~\bibnamefont {Utsumi}},\ }\href {https://doi.org/10.1103/PhysRevB.24.7385} {\bibfield  {journal} {\bibinfo  {journal} {Phys. Rev. B}\ }\textbf {\bibinfo {volume} {24}},\ \bibinfo {pages} {7385} (\bibinfo {year} {1981})}\BibitemShut {NoStop}%
\bibitem [{\citenamefont {Holas}(1987)}]{HOLAS1987}%
  \BibitemOpen
  \bibfield  {author} {\bibinfo {author} {\bibfnamefont {A.}~\bibnamefont {Holas}},\ }in\ \href {https://doi.org/10.1007/978-1-4613-1891-0} {\emph {\bibinfo {booktitle} {Strongly {{Coupled Plasma Physics}}}}}\ (\bibinfo  {publisher} {Plenum},\ \bibinfo {address} {New York},\ \bibinfo {year} {1987})\ p.\ \bibinfo {pages} {463}\BibitemShut {NoStop}%
\bibitem [{\citenamefont {Ortiz}\ and\ \citenamefont {Ballone}(1994)}]{ORTIZ1994}%
  \BibitemOpen
  \bibfield  {author} {\bibinfo {author} {\bibfnamefont {G.}~\bibnamefont {Ortiz}}\ and\ \bibinfo {author} {\bibfnamefont {P.}~\bibnamefont {Ballone}},\ }\href {https://doi.org/10.1103/PhysRevB.50.1391} {\bibfield  {journal} {\bibinfo  {journal} {Physical Review B}\ }\textbf {\bibinfo {volume} {50}},\ \bibinfo {pages} {1391} (\bibinfo {year} {1994})}\BibitemShut {NoStop}%
\bibitem [{\citenamefont {Chihara}\ and\ \citenamefont {Sasaki}(1979)}]{CHIHARA1979}%
  \BibitemOpen
  \bibfield  {author} {\bibinfo {author} {\bibfnamefont {J.}~\bibnamefont {Chihara}}\ and\ \bibinfo {author} {\bibfnamefont {K.}~\bibnamefont {Sasaki}},\ }\href {https://doi.org/10.1143/PTP.62.1533} {\bibfield  {journal} {\bibinfo  {journal} {Progress of Theoretical Physics}\ }\textbf {\bibinfo {volume} {62}},\ \bibinfo {pages} {1533} (\bibinfo {year} {1979})},\ \Eprint {https://arxiv.org/abs/https://academic.oup.com/ptp/article-pdf/62/6/1533/5397969/62-6-1533.pdf} {https://academic.oup.com/ptp/article-pdf/62/6/1533/5397969/62-6-1533.pdf} \BibitemShut {NoStop}%
\bibitem [{\citenamefont {Mahan}(1981)}]{MAHAN1981}%
  \BibitemOpen
  \bibfield  {author} {\bibinfo {author} {\bibfnamefont {G.~D.}\ \bibnamefont {Mahan}},\ }\href@noop {} {\emph {\bibinfo {title} {Many-particle physics}}}\ (\bibinfo  {publisher} {Plenum Press},\ \bibinfo {address} {New York},\ \bibinfo {year} {1981})\BibitemShut {NoStop}%
\bibitem [{\citenamefont {Sugiyama}\ \emph {et~al.}(1992)\citenamefont {Sugiyama}, \citenamefont {Bowen},\ and\ \citenamefont {Alder}}]{SUGIYAMA1992}%
  \BibitemOpen
  \bibfield  {author} {\bibinfo {author} {\bibfnamefont {G.}~\bibnamefont {Sugiyama}}, \bibinfo {author} {\bibfnamefont {C.}~\bibnamefont {Bowen}},\ and\ \bibinfo {author} {\bibfnamefont {B.~J.}\ \bibnamefont {Alder}},\ }\href {https://doi.org/10.1103/PhysRevB.46.13042} {\bibfield  {journal} {\bibinfo  {journal} {Physical Review B}\ }\textbf {\bibinfo {volume} {46}},\ \bibinfo {pages} {13042} (\bibinfo {year} {1992})}\BibitemShut {NoStop}%
\bibitem [{\citenamefont {Moroni}\ \emph {et~al.}(1992)\citenamefont {Moroni}, \citenamefont {Ceperley},\ and\ \citenamefont {Senatore}}]{MORONI1992}%
  \BibitemOpen
  \bibfield  {author} {\bibinfo {author} {\bibfnamefont {S.}~\bibnamefont {Moroni}}, \bibinfo {author} {\bibfnamefont {D.~M.}\ \bibnamefont {Ceperley}},\ and\ \bibinfo {author} {\bibfnamefont {G.}~\bibnamefont {Senatore}},\ }\href {https://doi.org/10.1103/PhysRevLett.69.1837} {\bibfield  {journal} {\bibinfo  {journal} {Physical Review Letters}\ }\textbf {\bibinfo {volume} {69}},\ \bibinfo {pages} {1837} (\bibinfo {year} {1992})}\BibitemShut {NoStop}%
\bibitem [{\citenamefont {Bowen}\ \emph {et~al.}(1994)\citenamefont {Bowen}, \citenamefont {Sugiyama},\ and\ \citenamefont {Alder}}]{BOWEN1994}%
  \BibitemOpen
  \bibfield  {author} {\bibinfo {author} {\bibfnamefont {C.}~\bibnamefont {Bowen}}, \bibinfo {author} {\bibfnamefont {G.}~\bibnamefont {Sugiyama}},\ and\ \bibinfo {author} {\bibfnamefont {B.~J.}\ \bibnamefont {Alder}},\ }\href {https://doi.org/10.1103/PhysRevB.50.14838} {\bibfield  {journal} {\bibinfo  {journal} {Physical Review B}\ }\textbf {\bibinfo {volume} {50}},\ \bibinfo {pages} {14838} (\bibinfo {year} {1994})}\BibitemShut {NoStop}%
\bibitem [{\citenamefont {Moroni}\ \emph {et~al.}(1995)\citenamefont {Moroni}, \citenamefont {Ceperley},\ and\ \citenamefont {Senatore}}]{MORONI1995}%
  \BibitemOpen
  \bibfield  {author} {\bibinfo {author} {\bibfnamefont {S.}~\bibnamefont {Moroni}}, \bibinfo {author} {\bibfnamefont {D.~M.}\ \bibnamefont {Ceperley}},\ and\ \bibinfo {author} {\bibfnamefont {G.}~\bibnamefont {Senatore}},\ }\href {https://doi.org/10.1103/PhysRevLett.75.689} {\bibfield  {journal} {\bibinfo  {journal} {Physical Review Letters}\ }\textbf {\bibinfo {volume} {75}},\ \bibinfo {pages} {689} (\bibinfo {year} {1995})}\BibitemShut {NoStop}%
\bibitem [{\citenamefont {Dornheim}\ \emph {et~al.}(2024)\citenamefont {Dornheim}, \citenamefont {Tolias}, \citenamefont {Vorberger},\ and\ \citenamefont {Moldabekov}}]{DORNHEIM2024}%
  \BibitemOpen
  \bibfield  {author} {\bibinfo {author} {\bibfnamefont {T.}~\bibnamefont {Dornheim}}, \bibinfo {author} {\bibfnamefont {P.}~\bibnamefont {Tolias}}, \bibinfo {author} {\bibfnamefont {J.}~\bibnamefont {Vorberger}},\ and\ \bibinfo {author} {\bibfnamefont {Z.~A.}\ \bibnamefont {Moldabekov}},\ }\href {https://doi.org/10.1209/0295-5075/ad5d88} {\bibfield  {journal} {\bibinfo  {journal} {Europhysics Letters}\ }\textbf {\bibinfo {volume} {147}},\ \bibinfo {pages} {36001} (\bibinfo {year} {2024})}\BibitemShut {NoStop}%
\bibitem [{\citenamefont {Dornheim}\ \emph {et~al.}(2020)\citenamefont {Dornheim}, \citenamefont {Sjostrom}, \citenamefont {Tanaka},\ and\ \citenamefont {Vorberger}}]{DORNHEIM2020}%
  \BibitemOpen
  \bibfield  {author} {\bibinfo {author} {\bibfnamefont {T.}~\bibnamefont {Dornheim}}, \bibinfo {author} {\bibfnamefont {T.}~\bibnamefont {Sjostrom}}, \bibinfo {author} {\bibfnamefont {S.}~\bibnamefont {Tanaka}},\ and\ \bibinfo {author} {\bibfnamefont {J.}~\bibnamefont {Vorberger}},\ }\href {https://doi.org/10.1103/PhysRevB.101.045129} {\bibfield  {journal} {\bibinfo  {journal} {Physical Review B}\ }\textbf {\bibinfo {volume} {101}},\ \bibinfo {pages} {045129} (\bibinfo {year} {2020})}\BibitemShut {NoStop}%
\bibitem [{\citenamefont {Dornheim}\ \emph {et~al.}(2019)\citenamefont {Dornheim}, \citenamefont {Vorberger}, \citenamefont {Groth}, \citenamefont {Hoffmann}, \citenamefont {Moldabekov},\ and\ \citenamefont {Bonitz}}]{DORNHEIM2019}%
  \BibitemOpen
  \bibfield  {author} {\bibinfo {author} {\bibfnamefont {T.}~\bibnamefont {Dornheim}}, \bibinfo {author} {\bibfnamefont {J.}~\bibnamefont {Vorberger}}, \bibinfo {author} {\bibfnamefont {S.}~\bibnamefont {Groth}}, \bibinfo {author} {\bibfnamefont {N.}~\bibnamefont {Hoffmann}}, \bibinfo {author} {\bibfnamefont {{\relax Zh}.~A.}\ \bibnamefont {Moldabekov}},\ and\ \bibinfo {author} {\bibfnamefont {M.}~\bibnamefont {Bonitz}},\ }\href {https://doi.org/10.1063/1.5123013} {\bibfield  {journal} {\bibinfo  {journal} {The Journal of Chemical Physics}\ }\textbf {\bibinfo {volume} {151}},\ \bibinfo {pages} {194104} (\bibinfo {year} {2019})}\BibitemShut {NoStop}%
\bibitem [{\citenamefont {Svensson}\ \emph {et~al.}(2026)\citenamefont {Svensson}, \citenamefont {Chuna}, \citenamefont {Vorberger}, \citenamefont {Moldabekov}, \citenamefont {Hamann}, \citenamefont {Schwalbe}, \citenamefont {Tolias},\ and\ \citenamefont {Dornheim}}]{SVENSSON2026}%
  \BibitemOpen
  \bibfield  {author} {\bibinfo {author} {\bibfnamefont {P.}~\bibnamefont {Svensson}}, \bibinfo {author} {\bibfnamefont {T.}~\bibnamefont {Chuna}}, \bibinfo {author} {\bibfnamefont {J.}~\bibnamefont {Vorberger}}, \bibinfo {author} {\bibfnamefont {Z.~A.}\ \bibnamefont {Moldabekov}}, \bibinfo {author} {\bibfnamefont {P.}~\bibnamefont {Hamann}}, \bibinfo {author} {\bibfnamefont {S.}~\bibnamefont {Schwalbe}}, \bibinfo {author} {\bibfnamefont {P.}~\bibnamefont {Tolias}},\ and\ \bibinfo {author} {\bibfnamefont {T.}~\bibnamefont {Dornheim}},\ }\href {https://doi.org/10.1063/5.0340968} {\bibfield  {journal} {\bibinfo  {journal} {The Journal of Chemical Physics}\ }\textbf {\bibinfo {volume} {165}},\ \bibinfo {pages} {044113} (\bibinfo {year} {2026})},\ \Eprint {https://arxiv.org/abs/2604.15934} {arXiv:2604.15934} \BibitemShut {NoStop}%
\bibitem [{\citenamefont {Corradini}\ \emph {et~al.}(1998)\citenamefont {Corradini}, \citenamefont {Del~Sole}, \citenamefont {Onida},\ and\ \citenamefont {Palummo}}]{CORRADINI1998}%
  \BibitemOpen
  \bibfield  {author} {\bibinfo {author} {\bibfnamefont {M.}~\bibnamefont {Corradini}}, \bibinfo {author} {\bibfnamefont {R.}~\bibnamefont {Del~Sole}}, \bibinfo {author} {\bibfnamefont {G.}~\bibnamefont {Onida}},\ and\ \bibinfo {author} {\bibfnamefont {M.}~\bibnamefont {Palummo}},\ }\href {https://doi.org/10.1103/PhysRevB.57.14569} {\bibfield  {journal} {\bibinfo  {journal} {Physical Review B}\ }\textbf {\bibinfo {volume} {57}},\ \bibinfo {pages} {14569} (\bibinfo {year} {1998})}\BibitemShut {NoStop}%
\bibitem [{\citenamefont {Olevano}\ \emph {et~al.}(1999)\citenamefont {Olevano}, \citenamefont {Palummo}, \citenamefont {Onida},\ and\ \citenamefont {Del~Sole}}]{OLEVANO1999}%
  \BibitemOpen
  \bibfield  {author} {\bibinfo {author} {\bibfnamefont {V.}~\bibnamefont {Olevano}}, \bibinfo {author} {\bibfnamefont {M.}~\bibnamefont {Palummo}}, \bibinfo {author} {\bibfnamefont {G.}~\bibnamefont {Onida}},\ and\ \bibinfo {author} {\bibfnamefont {R.}~\bibnamefont {Del~Sole}},\ }\href {https://doi.org/10.1103/PhysRevB.60.14224} {\bibfield  {journal} {\bibinfo  {journal} {Physical Review B}\ }\textbf {\bibinfo {volume} {60}},\ \bibinfo {pages} {14224} (\bibinfo {year} {1999})}\BibitemShut {NoStop}%
\bibitem [{\citenamefont {Palummo}\ \emph {et~al.}(1999)\citenamefont {Palummo}, \citenamefont {Onida}, \citenamefont {Del~Sole}, \citenamefont {Corradini},\ and\ \citenamefont {Reining}}]{PALUMMO1999}%
  \BibitemOpen
  \bibfield  {author} {\bibinfo {author} {\bibfnamefont {M.}~\bibnamefont {Palummo}}, \bibinfo {author} {\bibfnamefont {G.}~\bibnamefont {Onida}}, \bibinfo {author} {\bibfnamefont {R.}~\bibnamefont {Del~Sole}}, \bibinfo {author} {\bibfnamefont {M.}~\bibnamefont {Corradini}},\ and\ \bibinfo {author} {\bibfnamefont {L.}~\bibnamefont {Reining}},\ }\href {https://doi.org/10.1103/PhysRevB.60.11329} {\bibfield  {journal} {\bibinfo  {journal} {Physical Review B}\ }\textbf {\bibinfo {volume} {60}},\ \bibinfo {pages} {11329} (\bibinfo {year} {1999})},\ \Eprint {https://arxiv.org/abs/cond-mat/9901046} {cond-mat/9901046} \BibitemShut {NoStop}%
\bibitem [{\citenamefont {Cazzaniga}\ \emph {et~al.}(2011)\citenamefont {Cazzaniga}, \citenamefont {Weissker}, \citenamefont {Huotari}, \citenamefont {Pylkk\"anen}, \citenamefont {Salvestrini}, \citenamefont {Monaco}, \citenamefont {Onida},\ and\ \citenamefont {Reining}}]{CAZZANIGA2011}%
  \BibitemOpen
  \bibfield  {author} {\bibinfo {author} {\bibfnamefont {M.}~\bibnamefont {Cazzaniga}}, \bibinfo {author} {\bibfnamefont {H.-C.}\ \bibnamefont {Weissker}}, \bibinfo {author} {\bibfnamefont {S.}~\bibnamefont {Huotari}}, \bibinfo {author} {\bibfnamefont {T.}~\bibnamefont {Pylkk\"anen}}, \bibinfo {author} {\bibfnamefont {P.}~\bibnamefont {Salvestrini}}, \bibinfo {author} {\bibfnamefont {G.}~\bibnamefont {Monaco}}, \bibinfo {author} {\bibfnamefont {G.}~\bibnamefont {Onida}},\ and\ \bibinfo {author} {\bibfnamefont {L.}~\bibnamefont {Reining}},\ }\href {https://doi.org/10.1103/PhysRevB.84.075109} {\bibfield  {journal} {\bibinfo  {journal} {Phys. Rev. B}\ }\textbf {\bibinfo {volume} {84}},\ \bibinfo {pages} {075109} (\bibinfo {year} {2011})}\BibitemShut {NoStop}%
\bibitem [{\citenamefont {Ramakrishna}\ \emph {et~al.}(2021)\citenamefont {Ramakrishna}, \citenamefont {Cangi}, \citenamefont {Dornheim}, \citenamefont {Baczewski},\ and\ \citenamefont {Vorberger}}]{RAMAKRISHNA2021}%
  \BibitemOpen
  \bibfield  {author} {\bibinfo {author} {\bibfnamefont {K.}~\bibnamefont {Ramakrishna}}, \bibinfo {author} {\bibfnamefont {A.}~\bibnamefont {Cangi}}, \bibinfo {author} {\bibfnamefont {T.}~\bibnamefont {Dornheim}}, \bibinfo {author} {\bibfnamefont {A.}~\bibnamefont {Baczewski}},\ and\ \bibinfo {author} {\bibfnamefont {J.}~\bibnamefont {Vorberger}},\ }\href {https://doi.org/10.1103/PhysRevB.103.125118} {\bibfield  {journal} {\bibinfo  {journal} {Physical Review B}\ }\textbf {\bibinfo {volume} {103}},\ \bibinfo {pages} {125118} (\bibinfo {year} {2021})},\ \Eprint {https://arxiv.org/abs/2009.12163} {2009.12163} \BibitemShut {NoStop}%
\bibitem [{\citenamefont {Aouina}\ \emph {et~al.}(2025)\citenamefont {Aouina}, \citenamefont {Gatti},\ and\ \citenamefont {Reining}}]{AOUINA2025}%
  \BibitemOpen
  \bibfield  {author} {\bibinfo {author} {\bibfnamefont {A.}~\bibnamefont {Aouina}}, \bibinfo {author} {\bibfnamefont {M.}~\bibnamefont {Gatti}},\ and\ \bibinfo {author} {\bibfnamefont {L.}~\bibnamefont {Reining}},\ }\href {https://doi.org/10.1038/s41524-025-01712-4} {\bibfield  {journal} {\bibinfo  {journal} {npj Computational Materials}\ }\textbf {\bibinfo {volume} {11}},\ \bibinfo {pages} {242} (\bibinfo {year} {2025})}\BibitemShut {NoStop}%
\bibitem [{\citenamefont {Kaplan}\ and\ \citenamefont {Kukkonen}(2023)}]{KAPLAN2023}%
  \BibitemOpen
  \bibfield  {author} {\bibinfo {author} {\bibfnamefont {A.~D.}\ \bibnamefont {Kaplan}}\ and\ \bibinfo {author} {\bibfnamefont {C.~A.}\ \bibnamefont {Kukkonen}},\ }\href {https://doi.org/10.1103/PhysRevB.107.L201120} {\bibfield  {journal} {\bibinfo  {journal} {Physical Review B}\ }\textbf {\bibinfo {volume} {107}},\ \bibinfo {pages} {L201120} (\bibinfo {year} {2023})}\BibitemShut {NoStop}%
\bibitem [{\citenamefont {Kim}\ \emph {et~al.}(2018)\citenamefont {Kim} \emph {et~al.}}]{QMCPACK_1}%
  \BibitemOpen
  \bibfield  {author} {\bibinfo {author} {\bibfnamefont {J.}~\bibnamefont {Kim}} \emph {et~al.},\ }\href@noop {} {\bibfield  {journal} {\bibinfo  {journal} {Journal of Physics: Condensed Matter}\ }\textbf {\bibinfo {volume} {30}},\ \bibinfo {pages} {195901} (\bibinfo {year} {2018})}\BibitemShut {NoStop}%
\bibitem [{\citenamefont {Kent}\ \emph {et~al.}(2020)\citenamefont {Kent} \emph {et~al.}}]{QMCPACK_2}%
  \BibitemOpen
  \bibfield  {author} {\bibinfo {author} {\bibfnamefont {P.~R.~C.}\ \bibnamefont {Kent}} \emph {et~al.},\ }\href@noop {} {\bibfield  {journal} {\bibinfo  {journal} {The Journal of Chemical Physics}\ }\textbf {\bibinfo {volume} {152}},\ \bibinfo {pages} {174105} (\bibinfo {year} {2020})}\BibitemShut {NoStop}%
\bibitem [{\citenamefont {G{\"u}ne{\c{s}}}\ \emph {et~al.}(2026)\citenamefont {G{\"u}ne{\c{s}}}, \citenamefont {Yang}, \citenamefont {Gorelov}, \citenamefont {Morales}, \citenamefont {Gatti}, \citenamefont {Reining},\ and\ \citenamefont {Zhang}}]{GUNESprl}%
  \BibitemOpen
  \bibfield  {author} {\bibinfo {author} {\bibfnamefont {M.~H.}\ \bibnamefont {G{\"u}ne{\c{s}}}}, \bibinfo {author} {\bibfnamefont {Y.}~\bibnamefont {Yang}}, \bibinfo {author} {\bibfnamefont {V.}~\bibnamefont {Gorelov}}, \bibinfo {author} {\bibfnamefont {M.~A.}\ \bibnamefont {Morales}}, \bibinfo {author} {\bibfnamefont {M.}~\bibnamefont {Gatti}}, \bibinfo {author} {\bibfnamefont {L.}~\bibnamefont {Reining}},\ and\ \bibinfo {author} {\bibfnamefont {S.}~\bibnamefont {Zhang}}} (\bibinfo {year} {2026}),\ \bibinfo {note} {companion paper, submitted to Phys. Rev. Lett.}\BibitemShut {Stop}%
\bibitem [{\citenamefont {Ceperley}(1978)}]{CEPERLEY1978}%
  \BibitemOpen
  \bibfield  {author} {\bibinfo {author} {\bibfnamefont {D.}~\bibnamefont {Ceperley}},\ }\href {https://doi.org/10.1103/PhysRevB.18.3126} {\bibfield  {journal} {\bibinfo  {journal} {Physical Review B}\ }\textbf {\bibinfo {volume} {18}},\ \bibinfo {pages} {3126} (\bibinfo {year} {1978})}\BibitemShut {NoStop}%
\bibitem [{\citenamefont {Ceperley}\ and\ \citenamefont {Alder}(1980)}]{CEPERLEY1980}%
  \BibitemOpen
  \bibfield  {author} {\bibinfo {author} {\bibfnamefont {D.~M.}\ \bibnamefont {Ceperley}}\ and\ \bibinfo {author} {\bibfnamefont {B.~J.}\ \bibnamefont {Alder}},\ }\href {https://doi.org/10.1103/PhysRevLett.45.566} {\bibfield  {journal} {\bibinfo  {journal} {Physical Review Letters}\ }\textbf {\bibinfo {volume} {45}},\ \bibinfo {pages} {566} (\bibinfo {year} {1980})}\BibitemShut {NoStop}%
\bibitem [{\citenamefont {Tanatar}\ and\ \citenamefont {Ceperley}(1989)}]{TANATAR1989}%
  \BibitemOpen
  \bibfield  {author} {\bibinfo {author} {\bibfnamefont {B.}~\bibnamefont {Tanatar}}\ and\ \bibinfo {author} {\bibfnamefont {D.~M.}\ \bibnamefont {Ceperley}},\ }\href {https://doi.org/10.1103/PhysRevB.39.5005} {\bibfield  {journal} {\bibinfo  {journal} {Physical Review B}\ }\textbf {\bibinfo {volume} {39}},\ \bibinfo {pages} {5005} (\bibinfo {year} {1989})}\BibitemShut {NoStop}%
\bibitem [{\citenamefont {Jastrow}(1955)}]{JASTROW1955}%
  \BibitemOpen
  \bibfield  {author} {\bibinfo {author} {\bibfnamefont {R.}~\bibnamefont {Jastrow}},\ }\href {https://doi.org/10.1103/PhysRev.98.1479} {\bibfield  {journal} {\bibinfo  {journal} {Physical Review}\ }\textbf {\bibinfo {volume} {98}},\ \bibinfo {pages} {1479} (\bibinfo {year} {1955})}\BibitemShut {NoStop}%
\bibitem [{\citenamefont {Lee}\ \emph {et~al.}(1981)\citenamefont {Lee}, \citenamefont {Schmidt}, \citenamefont {Kalos},\ and\ \citenamefont {Chester}}]{LEE1981}%
  \BibitemOpen
  \bibfield  {author} {\bibinfo {author} {\bibfnamefont {M.~A.}\ \bibnamefont {Lee}}, \bibinfo {author} {\bibfnamefont {K.~E.}\ \bibnamefont {Schmidt}}, \bibinfo {author} {\bibfnamefont {M.~H.}\ \bibnamefont {Kalos}},\ and\ \bibinfo {author} {\bibfnamefont {G.~V.}\ \bibnamefont {Chester}},\ }\href {https://doi.org/10.1103/PhysRevLett.46.728} {\bibfield  {journal} {\bibinfo  {journal} {Physical Review Letters}\ }\textbf {\bibinfo {volume} {46}},\ \bibinfo {pages} {728} (\bibinfo {year} {1981})}\BibitemShut {NoStop}%
\bibitem [{\citenamefont {Schmidt}\ \emph {et~al.}(1981)\citenamefont {Schmidt}, \citenamefont {Lee}, \citenamefont {Kalos},\ and\ \citenamefont {Chester}}]{SCHMIDT1981}%
  \BibitemOpen
  \bibfield  {author} {\bibinfo {author} {\bibfnamefont {K.~E.}\ \bibnamefont {Schmidt}}, \bibinfo {author} {\bibfnamefont {M.~A.}\ \bibnamefont {Lee}}, \bibinfo {author} {\bibfnamefont {M.~H.}\ \bibnamefont {Kalos}},\ and\ \bibinfo {author} {\bibfnamefont {G.~V.}\ \bibnamefont {Chester}},\ }\href {https://doi.org/10.1103/PhysRevLett.47.807} {\bibfield  {journal} {\bibinfo  {journal} {Physical Review Letters}\ }\textbf {\bibinfo {volume} {47}},\ \bibinfo {pages} {807} (\bibinfo {year} {1981})}\BibitemShut {NoStop}%
\bibitem [{\citenamefont {Kwon}\ \emph {et~al.}(1993)\citenamefont {Kwon}, \citenamefont {Ceperley},\ and\ \citenamefont {Martin}}]{KWON1993}%
  \BibitemOpen
  \bibfield  {author} {\bibinfo {author} {\bibfnamefont {Y.}~\bibnamefont {Kwon}}, \bibinfo {author} {\bibfnamefont {D.~M.}\ \bibnamefont {Ceperley}},\ and\ \bibinfo {author} {\bibfnamefont {R.~M.}\ \bibnamefont {Martin}},\ }\href {https://doi.org/10.1103/PhysRevB.48.12037} {\bibfield  {journal} {\bibinfo  {journal} {Physical Review B}\ }\textbf {\bibinfo {volume} {48}},\ \bibinfo {pages} {12037} (\bibinfo {year} {1993})}\BibitemShut {NoStop}%
\bibitem [{\citenamefont {Kwon}\ \emph {et~al.}(1998)\citenamefont {Kwon}, \citenamefont {Ceperley},\ and\ \citenamefont {Martin}}]{KWON1998}%
  \BibitemOpen
  \bibfield  {author} {\bibinfo {author} {\bibfnamefont {Y.}~\bibnamefont {Kwon}}, \bibinfo {author} {\bibfnamefont {D.~M.}\ \bibnamefont {Ceperley}},\ and\ \bibinfo {author} {\bibfnamefont {R.~M.}\ \bibnamefont {Martin}},\ }\href {https://doi.org/10.1103/PhysRevB.58.6800} {\bibfield  {journal} {\bibinfo  {journal} {Physical Review B}\ }\textbf {\bibinfo {volume} {58}},\ \bibinfo {pages} {6800} (\bibinfo {year} {1998})}\BibitemShut {NoStop}%
\bibitem [{\citenamefont {Giannozzi}\ \emph {et~al.}(2009)\citenamefont {Giannozzi} \emph {et~al.}}]{q-e_short}%
  \BibitemOpen
  \bibfield  {author} {\bibinfo {author} {\bibfnamefont {P.}~\bibnamefont {Giannozzi}} \emph {et~al.},\ }\href {https://doi.org/10.1088/0953-8984/21/39/395502} {\bibfield  {journal} {\bibinfo  {journal} {Journal of Physics: Condensed Matter}\ }\textbf {\bibinfo {volume} {21}},\ \bibinfo {pages} {395502} (\bibinfo {year} {2009})},\ \Eprint {https://arxiv.org/abs/0906.2569} {arXiv:0906.2569 [cond-mat]} \BibitemShut {NoStop}%
\bibitem [{\citenamefont {Azadi}\ and\ \citenamefont {Drummond}(2022)}]{AZADI2022}%
  \BibitemOpen
  \bibfield  {author} {\bibinfo {author} {\bibfnamefont {S.}~\bibnamefont {Azadi}}\ and\ \bibinfo {author} {\bibfnamefont {N.~D.}\ \bibnamefont {Drummond}},\ }\href {https://doi.org/10.1103/PhysRevB.105.245135} {\bibfield  {journal} {\bibinfo  {journal} {Physical Review B}\ }\textbf {\bibinfo {volume} {105}},\ \bibinfo {pages} {245135} (\bibinfo {year} {2022})}\BibitemShut {NoStop}%
\bibitem [{\citenamefont {Drummond}\ \emph {et~al.}(2004)\citenamefont {Drummond}, \citenamefont {Radnai}, \citenamefont {Trail}, \citenamefont {Towler},\ and\ \citenamefont {Needs}}]{DRUMMOND2004}%
  \BibitemOpen
  \bibfield  {author} {\bibinfo {author} {\bibfnamefont {N.~D.}\ \bibnamefont {Drummond}}, \bibinfo {author} {\bibfnamefont {Z.}~\bibnamefont {Radnai}}, \bibinfo {author} {\bibfnamefont {J.~R.}\ \bibnamefont {Trail}}, \bibinfo {author} {\bibfnamefont {M.~D.}\ \bibnamefont {Towler}},\ and\ \bibinfo {author} {\bibfnamefont {R.~J.}\ \bibnamefont {Needs}},\ }\href {https://doi.org/10.1103/PhysRevB.69.085116} {\bibfield  {journal} {\bibinfo  {journal} {Phys. Rev. B}\ }\textbf {\bibinfo {volume} {69}},\ \bibinfo {pages} {085116} (\bibinfo {year} {2004})}\BibitemShut {NoStop}%
\bibitem [{\citenamefont {Holzmann}\ and\ \citenamefont {Moroni}(2020)}]{HOLZMANN2020}%
  \BibitemOpen
  \bibfield  {author} {\bibinfo {author} {\bibfnamefont {M.}~\bibnamefont {Holzmann}}\ and\ \bibinfo {author} {\bibfnamefont {S.}~\bibnamefont {Moroni}},\ }\href {https://doi.org/10.1103/PhysRevLett.124.206404} {\bibfield  {journal} {\bibinfo  {journal} {Phys. Rev. Lett.}\ }\textbf {\bibinfo {volume} {124}},\ \bibinfo {pages} {206404} (\bibinfo {year} {2020})}\BibitemShut {NoStop}%
\bibitem [{\citenamefont {Nepal}\ \emph {et~al.}(2020)\citenamefont {Nepal}, \citenamefont {Adhikari}, \citenamefont {Neupane}, \citenamefont {Ruan}, \citenamefont {Neupane},\ and\ \citenamefont {Ruzsinszky}}]{NEPAL2020}%
  \BibitemOpen
  \bibfield  {author} {\bibinfo {author} {\bibfnamefont {N.~K.}\ \bibnamefont {Nepal}}, \bibinfo {author} {\bibfnamefont {S.}~\bibnamefont {Adhikari}}, \bibinfo {author} {\bibfnamefont {B.}~\bibnamefont {Neupane}}, \bibinfo {author} {\bibfnamefont {S.}~\bibnamefont {Ruan}}, \bibinfo {author} {\bibfnamefont {S.}~\bibnamefont {Neupane}},\ and\ \bibinfo {author} {\bibfnamefont {A.}~\bibnamefont {Ruzsinszky}},\ }\href {https://doi.org/10.1103/PhysRevB.101.195137} {\bibfield  {journal} {\bibinfo  {journal} {Phys. Rev. B}\ }\textbf {\bibinfo {volume} {101}},\ \bibinfo {pages} {195137} (\bibinfo {year} {2020})}\BibitemShut {NoStop}%
\bibitem [{\citenamefont {Pellegrini}\ \emph {et~al.}(2023)\citenamefont {Pellegrini}, \citenamefont {Kukkonen},\ and\ \citenamefont {Sanna}}]{PELLEGRINI2023}%
  \BibitemOpen
  \bibfield  {author} {\bibinfo {author} {\bibfnamefont {C.}~\bibnamefont {Pellegrini}}, \bibinfo {author} {\bibfnamefont {C.}~\bibnamefont {Kukkonen}},\ and\ \bibinfo {author} {\bibfnamefont {A.}~\bibnamefont {Sanna}},\ }\href {https://doi.org/10.1103/PhysRevB.108.064511} {\bibfield  {journal} {\bibinfo  {journal} {Phys. Rev. B}\ }\textbf {\bibinfo {volume} {108}},\ \bibinfo {pages} {064511} (\bibinfo {year} {2023})}\BibitemShut {NoStop}%
\bibitem [{\citenamefont {Li}\ \emph {et~al.}(2026)\citenamefont {Li}, \citenamefont {Hou}, \citenamefont {Wang}, \citenamefont {Deng},\ and\ \citenamefont {Chen}}]{LI2026}%
  \BibitemOpen
  \bibfield  {author} {\bibinfo {author} {\bibfnamefont {Z.}~\bibnamefont {Li}}, \bibinfo {author} {\bibfnamefont {P.}~\bibnamefont {Hou}}, \bibinfo {author} {\bibfnamefont {B.-Z.}\ \bibnamefont {Wang}}, \bibinfo {author} {\bibfnamefont {Y.}~\bibnamefont {Deng}},\ and\ \bibinfo {author} {\bibfnamefont {K.}~\bibnamefont {Chen}},\ }\href {https://doi.org/10.1088/0256-307X/43/8/080705} {\bibfield  {journal} {\bibinfo  {journal} {Chinese Physics Letters}\ }\textbf {\bibinfo {volume} {43}},\ \bibinfo {pages} {080705} (\bibinfo {year} {2026})},\ \Eprint {https://arxiv.org/abs/2511.22927} {arXiv:2511.22927} \BibitemShut {NoStop}%
\bibitem [{Note1()}]{Note1}%
  \BibitemOpen
  \bibinfo {note} {In fact, the same behavior is also reported in Ref.~\cite {CORRADINI1998}, where the authors report the unpublished $r_s=100$ data of Moroni \protect \textit {et al.} for a fully polarized gas.}\BibitemShut {Stop}%
\bibitem [{\citenamefont {Mithen}\ \emph {et~al.}(2012)\citenamefont {Mithen}, \citenamefont {Daligault},\ and\ \citenamefont {Gregori}}]{MITHEN2012}%
  \BibitemOpen
  \bibfield  {author} {\bibinfo {author} {\bibfnamefont {J.~P.}\ \bibnamefont {Mithen}}, \bibinfo {author} {\bibfnamefont {J.}~\bibnamefont {Daligault}},\ and\ \bibinfo {author} {\bibfnamefont {G.}~\bibnamefont {Gregori}},\ }\href {https://doi.org/10.1103/PhysRevE.85.056407} {\bibfield  {journal} {\bibinfo  {journal} {Phys. Rev. E}\ }\textbf {\bibinfo {volume} {85}},\ \bibinfo {pages} {056407} (\bibinfo {year} {2012})}\BibitemShut {NoStop}%
\bibitem [{\citenamefont {Senatore}\ \emph {et~al.}(1999)\citenamefont {Senatore}, \citenamefont {Moroni},\ and\ \citenamefont {Ceperley}}]{SENATORE1999}%
  \BibitemOpen
  \bibfield  {author} {\bibinfo {author} {\bibfnamefont {G.}~\bibnamefont {Senatore}}, \bibinfo {author} {\bibfnamefont {S.}~\bibnamefont {Moroni}},\ and\ \bibinfo {author} {\bibfnamefont {D.~M.}\ \bibnamefont {Ceperley}},\ }in\ \href@noop {} {\emph {\bibinfo {booktitle} {Quantum {{Monte Carlo Methods}} in {{Physics}} and {{Chemistry}}}}},\ \bibinfo {series} {NATO {{Science Series C}}: {{Mathematical}} and {{Physical Sciences}}}, Vol.\ \bibinfo {volume} {525},\ \bibinfo {editor} {edited by\ \bibinfo {editor} {\bibfnamefont {M.~P.}\ \bibnamefont {Nightingale}}\ and\ \bibinfo {editor} {\bibfnamefont {C.~J.}\ \bibnamefont {Umrigar}}}\ (\bibinfo  {publisher} {Kluwer Academic},\ \bibinfo {address} {Dordrecht},\ \bibinfo {year} {1999})\ pp.\ \bibinfo {pages} {183--212}\BibitemShut {NoStop}%
\bibitem [{\citenamefont {Dharma-wardana}\ and\ \citenamefont {Perrot}(2003)}]{WARDANA2003}%
  \BibitemOpen
  \bibfield  {author} {\bibinfo {author} {\bibfnamefont {M.~W.~C.}\ \bibnamefont {Dharma-wardana}}\ and\ \bibinfo {author} {\bibfnamefont {F.}~\bibnamefont {Perrot}},\ }\href {https://doi.org/10.1209/epl/i2003-00593-6} {\bibfield  {journal} {\bibinfo  {journal} {Europhysics Letters}\ }\textbf {\bibinfo {volume} {63}},\ \bibinfo {pages} {660} (\bibinfo {year} {2003})}\BibitemShut {NoStop}%
\bibitem [{\citenamefont {Dornheim}\ \emph {et~al.}(2018)\citenamefont {Dornheim}, \citenamefont {Groth},\ and\ \citenamefont {Bonitz}}]{DORNHEIM2018}%
  \BibitemOpen
  \bibfield  {author} {\bibinfo {author} {\bibfnamefont {T.}~\bibnamefont {Dornheim}}, \bibinfo {author} {\bibfnamefont {S.}~\bibnamefont {Groth}},\ and\ \bibinfo {author} {\bibfnamefont {M.}~\bibnamefont {Bonitz}},\ }\href {https://doi.org/10.1016/j.physrep.2018.04.001} {\bibfield  {journal} {\bibinfo  {journal} {Physics Reports}\ }\textbf {\bibinfo {volume} {744}},\ \bibinfo {pages} {1} (\bibinfo {year} {2018})}\BibitemShut {NoStop}%
\bibitem [{Note2()}]{Note2}%
  \BibitemOpen
  \bibinfo {note} {At $r_s = 100$ (not shown), close to Wigner crystallization, the PIMC data for $G(q)$ no longer agree quantitatively, although the same structure is present.}\BibitemShut {Stop}%
\bibitem [{Note3()}]{Note3}%
  \BibitemOpen
  \bibinfo {note} {The height of this peak increases with $r_s$ and then saturates, with the $r_s=80$ and $90$ curves almost on top of each other, i.e., the local order saturates before long-range order sets in.}\BibitemShut {Stop}%
\bibitem [{\citenamefont {Davoudi}\ \emph {et~al.}(2003)\citenamefont {Davoudi}, \citenamefont {Asgari}, \citenamefont {Polini},\ and\ \citenamefont {Tosi}}]{DAVOUDI2003}%
  \BibitemOpen
  \bibfield  {author} {\bibinfo {author} {\bibfnamefont {B.}~\bibnamefont {Davoudi}}, \bibinfo {author} {\bibfnamefont {R.}~\bibnamefont {Asgari}}, \bibinfo {author} {\bibfnamefont {M.}~\bibnamefont {Polini}},\ and\ \bibinfo {author} {\bibfnamefont {M.~P.}\ \bibnamefont {Tosi}},\ }\href {https://doi.org/10.1103/PhysRevB.67.172503} {\bibfield  {journal} {\bibinfo  {journal} {Phys. Rev. B}\ }\textbf {\bibinfo {volume} {67}},\ \bibinfo {pages} {172503} (\bibinfo {year} {2003})}\BibitemShut {NoStop}%
\bibitem [{\citenamefont {Gattobigio}\ \emph {et~al.}(2005)\citenamefont {Gattobigio}, \citenamefont {Capuzzi}, \citenamefont {Polini}, \citenamefont {Asgari},\ and\ \citenamefont {Tosi}}]{GATTOBIGIO2005}%
  \BibitemOpen
  \bibfield  {author} {\bibinfo {author} {\bibfnamefont {M.}~\bibnamefont {Gattobigio}}, \bibinfo {author} {\bibfnamefont {P.}~\bibnamefont {Capuzzi}}, \bibinfo {author} {\bibfnamefont {M.}~\bibnamefont {Polini}}, \bibinfo {author} {\bibfnamefont {R.}~\bibnamefont {Asgari}},\ and\ \bibinfo {author} {\bibfnamefont {M.~P.}\ \bibnamefont {Tosi}},\ }\href {https://doi.org/10.1103/PhysRevB.72.045306} {\bibfield  {journal} {\bibinfo  {journal} {Phys. Rev. B}\ }\textbf {\bibinfo {volume} {72}},\ \bibinfo {pages} {045306} (\bibinfo {year} {2005})}\BibitemShut {NoStop}%
\bibitem [{\citenamefont {Smith}\ \emph {et~al.}(2024)\citenamefont {Smith}, \citenamefont {Chen}, \citenamefont {Levy}, \citenamefont {Yang}, \citenamefont {Morales},\ and\ \citenamefont {Zhang}}]{SMITH2024}%
  \BibitemOpen
  \bibfield  {author} {\bibinfo {author} {\bibfnamefont {C.}~\bibnamefont {Smith}}, \bibinfo {author} {\bibfnamefont {Y.}~\bibnamefont {Chen}}, \bibinfo {author} {\bibfnamefont {R.}~\bibnamefont {Levy}}, \bibinfo {author} {\bibfnamefont {Y.}~\bibnamefont {Yang}}, \bibinfo {author} {\bibfnamefont {M.~A.}\ \bibnamefont {Morales}},\ and\ \bibinfo {author} {\bibfnamefont {S.}~\bibnamefont {Zhang}},\ }\href {https://doi.org/10.1103/PhysRevLett.133.266504} {\bibfield  {journal} {\bibinfo  {journal} {Phys. Rev. Lett.}\ }\textbf {\bibinfo {volume} {133}},\ \bibinfo {pages} {266504} (\bibinfo {year} {2024})}\BibitemShut {NoStop}%
\bibitem [{\citenamefont {Pescia}\ \emph {et~al.}(2024)\citenamefont {Pescia}, \citenamefont {Nys}, \citenamefont {Kim}, \citenamefont {Lovato},\ and\ \citenamefont {Carleo}}]{PESCIA2024}%
  \BibitemOpen
  \bibfield  {author} {\bibinfo {author} {\bibfnamefont {G.}~\bibnamefont {Pescia}}, \bibinfo {author} {\bibfnamefont {J.}~\bibnamefont {Nys}}, \bibinfo {author} {\bibfnamefont {J.}~\bibnamefont {Kim}}, \bibinfo {author} {\bibfnamefont {A.}~\bibnamefont {Lovato}},\ and\ \bibinfo {author} {\bibfnamefont {G.}~\bibnamefont {Carleo}},\ }\href {https://doi.org/10.1103/PhysRevB.110.035108} {\bibfield  {journal} {\bibinfo  {journal} {Phys. Rev. B}\ }\textbf {\bibinfo {volume} {110}},\ \bibinfo {pages} {035108} (\bibinfo {year} {2024})}\BibitemShut {NoStop}%
\bibitem [{\citenamefont {Gori-Giorgi}\ and\ \citenamefont {Perdew}(2002)}]{GIORGI2002}%
  \BibitemOpen
  \bibfield  {author} {\bibinfo {author} {\bibfnamefont {P.}~\bibnamefont {Gori-Giorgi}}\ and\ \bibinfo {author} {\bibfnamefont {J.~P.}\ \bibnamefont {Perdew}},\ }\href {https://doi.org/10.1103/PhysRevB.66.165118} {\bibfield  {journal} {\bibinfo  {journal} {Phys. Rev. B}\ }\textbf {\bibinfo {volume} {66}},\ \bibinfo {pages} {165118} (\bibinfo {year} {2002})}\BibitemShut {NoStop}%
\bibitem [{\citenamefont {Ichimaru}(2004)}]{Ichimaru2004}%
  \BibitemOpen
  \bibfield  {author} {\bibinfo {author} {\bibfnamefont {S.}~\bibnamefont {Ichimaru}},\ }\href@noop {} {\emph {\bibinfo {title} {Statistical Plasma Physics, Volume {I}: Basic Principles}}}\ (\bibinfo  {publisher} {Westview Press},\ \bibinfo {address} {Boulder, CO},\ \bibinfo {year} {2004})\BibitemShut {NoStop}%
\bibitem [{\citenamefont {Runge}\ and\ \citenamefont {Gross}(1984)}]{Runge1984}%
  \BibitemOpen
  \bibfield  {author} {\bibinfo {author} {\bibfnamefont {E.}~\bibnamefont {Runge}}\ and\ \bibinfo {author} {\bibfnamefont {E.~K.~U.}\ \bibnamefont {Gross}},\ }\href {https://doi.org/10.1103/PhysRevLett.52.997} {\bibfield  {journal} {\bibinfo  {journal} {Phys. Rev. Lett.}\ }\textbf {\bibinfo {volume} {52}},\ \bibinfo {pages} {997} (\bibinfo {year} {1984})}\BibitemShut {NoStop}%
\bibitem [{\citenamefont {Petersilka}\ \emph {et~al.}(1996)\citenamefont {Petersilka}, \citenamefont {Gossmann},\ and\ \citenamefont {Gross}}]{Petersilka1996}%
  \BibitemOpen
  \bibfield  {author} {\bibinfo {author} {\bibfnamefont {M.}~\bibnamefont {Petersilka}}, \bibinfo {author} {\bibfnamefont {U.~J.}\ \bibnamefont {Gossmann}},\ and\ \bibinfo {author} {\bibfnamefont {E.~K.~U.}\ \bibnamefont {Gross}},\ }\href {https://doi.org/10.1103/PhysRevLett.76.1212} {\bibfield  {journal} {\bibinfo  {journal} {Phys. Rev. Lett.}\ }\textbf {\bibinfo {volume} {76}},\ \bibinfo {pages} {1212} (\bibinfo {year} {1996})}\BibitemShut {NoStop}%
\bibitem [{\citenamefont {Ullrich}(2012)}]{Ullrich2012}%
  \BibitemOpen
  \bibfield  {author} {\bibinfo {author} {\bibfnamefont {C.}~\bibnamefont {Ullrich}},\ }\href@noop {} {\emph {\bibinfo {title} {Time-Dependent Density-Functional Theory: Concepts and Applications}}},\ Oxford Graduate Texts\ (\bibinfo  {publisher} {OUP Oxford},\ \bibinfo {year} {2012})\BibitemShut {NoStop}%
\bibitem [{\citenamefont {Koskelo}\ \emph {et~al.}(2025)\citenamefont {Koskelo}, \citenamefont {Reining},\ and\ \citenamefont {Gatti}}]{KOSKELO2025}%
  \BibitemOpen
  \bibfield  {author} {\bibinfo {author} {\bibfnamefont {J.}~\bibnamefont {Koskelo}}, \bibinfo {author} {\bibfnamefont {L.}~\bibnamefont {Reining}},\ and\ \bibinfo {author} {\bibfnamefont {M.}~\bibnamefont {Gatti}},\ }\href {https://doi.org/10.1103/PhysRevLett.134.046402} {\bibfield  {journal} {\bibinfo  {journal} {Physical Review Letters}\ }\textbf {\bibinfo {volume} {134}},\ \bibinfo {pages} {046402} (\bibinfo {year} {2025})}\BibitemShut {NoStop}%
\bibitem [{\citenamefont {Chulkov}\ \emph {et~al.}(2006)\citenamefont {Chulkov}, \citenamefont {Borisov}, \citenamefont {Gauyacq}, \citenamefont {Sánchez-Portal}, \citenamefont {Silkin}, \citenamefont {Zhukov},\ and\ \citenamefont {Echenique}}]{Chulkov2006}%
  \BibitemOpen
  \bibfield  {author} {\bibinfo {author} {\bibfnamefont {E.~V.}\ \bibnamefont {Chulkov}}, \bibinfo {author} {\bibfnamefont {A.~G.}\ \bibnamefont {Borisov}}, \bibinfo {author} {\bibfnamefont {J.~P.}\ \bibnamefont {Gauyacq}}, \bibinfo {author} {\bibfnamefont {D.}~\bibnamefont {Sánchez-Portal}}, \bibinfo {author} {\bibfnamefont {V.~M.}\ \bibnamefont {Silkin}}, \bibinfo {author} {\bibfnamefont {V.~P.}\ \bibnamefont {Zhukov}},\ and\ \bibinfo {author} {\bibfnamefont {P.~M.}\ \bibnamefont {Echenique}},\ }\href {https://doi.org/10.1021/cr050166o} {\bibfield  {journal} {\bibinfo  {journal} {Chemical Reviews}\ }\textbf {\bibinfo {volume} {106}},\ \bibinfo {pages} {4160} (\bibinfo {year} {2006})}\BibitemShut {NoStop}%
\bibitem [{\citenamefont {Pitarke}\ and\ \citenamefont {Perdew}(2003)}]{Pitarke2003}%
  \BibitemOpen
  \bibfield  {author} {\bibinfo {author} {\bibfnamefont {J.~M.}\ \bibnamefont {Pitarke}}\ and\ \bibinfo {author} {\bibfnamefont {J.~P.}\ \bibnamefont {Perdew}},\ }\href {https://doi.org/10.1103/PhysRevB.67.045101} {\bibfield  {journal} {\bibinfo  {journal} {Phys. Rev. B}\ }\textbf {\bibinfo {volume} {67}},\ \bibinfo {pages} {045101} (\bibinfo {year} {2003})}\BibitemShut {NoStop}%
\bibitem [{\citenamefont {Patrick}\ and\ \citenamefont {Thygesen}(2015)}]{Patrick2015}%
  \BibitemOpen
  \bibfield  {author} {\bibinfo {author} {\bibfnamefont {C.~E.}\ \bibnamefont {Patrick}}\ and\ \bibinfo {author} {\bibfnamefont {K.~S.}\ \bibnamefont {Thygesen}},\ }\href {https://doi.org/10.1063/1.4919236} {\bibfield  {journal} {\bibinfo  {journal} {The Journal of Chemical Physics}\ }\textbf {\bibinfo {volume} {143}},\ \bibinfo {pages} {102802} (\bibinfo {year} {2015})}\BibitemShut {NoStop}%
\bibitem [{\citenamefont {Olsen}\ \emph {et~al.}(2019)\citenamefont {Olsen}, \citenamefont {Patrick}, \citenamefont {Bates}, \citenamefont {Ruzsinszky},\ and\ \citenamefont {Thygesen}}]{Olsen2019}%
  \BibitemOpen
  \bibfield  {author} {\bibinfo {author} {\bibfnamefont {T.}~\bibnamefont {Olsen}}, \bibinfo {author} {\bibfnamefont {C.~E.}\ \bibnamefont {Patrick}}, \bibinfo {author} {\bibfnamefont {J.~E.}\ \bibnamefont {Bates}}, \bibinfo {author} {\bibfnamefont {A.}~\bibnamefont {Ruzsinszky}},\ and\ \bibinfo {author} {\bibfnamefont {K.~S.}\ \bibnamefont {Thygesen}},\ }\href {https://doi.org/10.1038/s41524-019-0242-8} {\bibfield  {journal} {\bibinfo  {journal} {npj Computational Materials}\ }\textbf {\bibinfo {volume} {5}},\ \bibinfo {pages} {106} (\bibinfo {year} {2019})}\BibitemShut {NoStop}%
\bibitem [{\citenamefont {Takada}(1993)}]{TAKADA1993}%
  \BibitemOpen
  \bibfield  {author} {\bibinfo {author} {\bibfnamefont {Y.}~\bibnamefont {Takada}},\ }\href {https://doi.org/10.1103/PhysRevB.47.5202} {\bibfield  {journal} {\bibinfo  {journal} {Physical Review B}\ }\textbf {\bibinfo {volume} {47}},\ \bibinfo {pages} {5202} (\bibinfo {year} {1993})}\BibitemShut {NoStop}%
\bibitem [{\citenamefont {Dornheim}\ \emph {et~al.}(2022)\citenamefont {Dornheim}, \citenamefont {Tolias}, \citenamefont {Moldabekov}, \citenamefont {Cangi},\ and\ \citenamefont {Vorberger}}]{DORNHEIM2022b}%
  \BibitemOpen
  \bibfield  {author} {\bibinfo {author} {\bibfnamefont {T.}~\bibnamefont {Dornheim}}, \bibinfo {author} {\bibfnamefont {P.}~\bibnamefont {Tolias}}, \bibinfo {author} {\bibfnamefont {Z.~A.}\ \bibnamefont {Moldabekov}}, \bibinfo {author} {\bibfnamefont {A.}~\bibnamefont {Cangi}},\ and\ \bibinfo {author} {\bibfnamefont {J.}~\bibnamefont {Vorberger}},\ }\href {https://doi.org/10.1063/5.0097768} {\bibfield  {journal} {\bibinfo  {journal} {The Journal of Chemical Physics}\ }\textbf {\bibinfo {volume} {156}},\ \bibinfo {pages} {244113} (\bibinfo {year} {2022})}\BibitemShut {NoStop}%
\bibitem [{\citenamefont {G{\"u}ne{\c{s}}}(2026)}]{zenodo}%
  \BibitemOpen
  \bibfield  {author} {\bibinfo {author} {\bibfnamefont {M.~H.}\ \bibnamefont {G{\"u}ne{\c{s}}}},\ }\href@noop {} {\bibinfo {title} {{Static response, structure factor and pair correlation of the 3D electron gas from DMC}}},\ \bibinfo {howpublished} {Zenodo, version v1.0} (\bibinfo {year} {2026}),\ \bibinfo {note} {\href{https://doi.org/10.5281/zenodo.23436447}{doi:10.5281/zenodo.23436447}}\BibitemShut {NoStop}%
\bibitem [{\citenamefont {Perdew}\ and\ \citenamefont {Wang}(1992)}]{PERDEW1992}%
  \BibitemOpen
  \bibfield  {author} {\bibinfo {author} {\bibfnamefont {J.~P.}\ \bibnamefont {Perdew}}\ and\ \bibinfo {author} {\bibfnamefont {Y.}~\bibnamefont {Wang}},\ }\href {https://doi.org/10.1103/PhysRevB.45.13244} {\bibfield  {journal} {\bibinfo  {journal} {Phys. Rev. B}\ }\textbf {\bibinfo {volume} {45}},\ \bibinfo {pages} {13244} (\bibinfo {year} {1992})}\BibitemShut {NoStop}%
\bibitem [{\citenamefont {Perdew}\ and\ \citenamefont {Zunger}(1981)}]{PERDEW1981}%
  \BibitemOpen
  \bibfield  {author} {\bibinfo {author} {\bibfnamefont {J.~P.}\ \bibnamefont {Perdew}}\ and\ \bibinfo {author} {\bibfnamefont {A.}~\bibnamefont {Zunger}},\ }\href {https://doi.org/10.1103/PhysRevB.23.5048} {\bibfield  {journal} {\bibinfo  {journal} {Phys. Rev. B}\ }\textbf {\bibinfo {volume} {23}},\ \bibinfo {pages} {5048} (\bibinfo {year} {1981})}\BibitemShut {NoStop}%
\bibitem [{\citenamefont {Lin}\ \emph {et~al.}(2001)\citenamefont {Lin}, \citenamefont {Zong},\ and\ \citenamefont {Ceperley}}]{Lin_twist-averaged-boundary_2001}%
  \BibitemOpen
  \bibfield  {author} {\bibinfo {author} {\bibfnamefont {C.}~\bibnamefont {Lin}}, \bibinfo {author} {\bibfnamefont {F.~H.}\ \bibnamefont {Zong}},\ and\ \bibinfo {author} {\bibfnamefont {D.~M.}\ \bibnamefont {Ceperley}},\ }\href {https://doi.org/10.1103/PhysRevE.64.016702} {\bibfield  {journal} {\bibinfo  {journal} {Phys. Rev. E}\ }\textbf {\bibinfo {volume} {64}},\ \bibinfo {pages} {016702} (\bibinfo {year} {2001})}\BibitemShut {NoStop}%
\end{thebibliography}%

\end{document}